\documentclass[%
reprint,
superscriptaddress,
showpacs,
 amsmath,amssymb,
 aps,
floatfix,
]{revtex4-1}

\usepackage{graphicx}
\usepackage{dcolumn}
\usepackage{bm}
\usepackage{xcolor}
\usepackage{color}
\usepackage{float}
\usepackage[english]{babel}
\usepackage{tgtermes}
\usepackage{mathptmx}
\usepackage[T1]{fontenc}
\usepackage{exscale}

\newcommand{\be}{\begin{equation}}
\newcommand{\ee}{\end{equation}}

\begin{document}

\preprint{APS/123-QED}

\title{Atomic structure of grain boundaries in iron modeled using the atomic density function}

\author{O. Kapikranian}
\email{oleksandr.kapikranyan@univ-rouen.fr,  akap@icmp.lviv.ua}
\affiliation{GPM UMR 6634 CNRS, Universit\'e de Rouen, F-76801 Saint Etienne du Rouvray, France}
\affiliation{Institute for condensed matter physics of the National Acad. of Sc. of Ukraine, 79011 Lviv, Ukraine}
\author{H. Zapolsky}
\affiliation{GPM UMR 6634 CNRS, Universit\'e de Rouen, F-76801 Saint Etienne du Rouvray, France}
\author{C. Domain}
\affiliation{D\'epartement MMC, E.D.F. - R\&D, Les Renardi\`eres, F-77250 Moret-sur-Loing, France}
\author{R. Patte}
\affiliation{GPM UMR 6634 CNRS, Universit\'e de Rouen, F-76801 Saint Etienne du Rouvray, France}
\author{C. Pareige}
\affiliation{GPM UMR 6634 CNRS, Universit\'e de Rouen, F-76801 Saint Etienne du Rouvray, France}
\author{B. Radiguet}
\affiliation{GPM UMR 6634 CNRS, Universit\'e de Rouen, F-76801 Saint Etienne du Rouvray, France}
\author{P. Pareige}
\affiliation{GPM UMR 6634 CNRS, Universit\'e de Rouen, F-76801 Saint Etienne du Rouvray, France}

\date{\today}

\begin{abstract}
A model based on the continuous atomic density function (ADF) approach is applied to predict the atomic structure of grain boundaries (GBs) in iron. Symmetrical $[100]$ and $[110]$ tilt GBs in bcc iron are modeled with the ADF method and relaxed afterwards in molecular dynamics (MD) simulations. The shape of the GB energy curve obtained in the ADF model reproduces well the peculiarities of the angles of $70.53^\circ$ ($\Sigma 3 (112)$) and $129.52^\circ$ ($\Sigma 11 (332)$) for $[110]$ tilt GBs. The results of MD relaxation with an embedded-atom method potential for iron confirm that the atomic GB configurations obtained in ADF modeling are very close to equilibrium ones. The developed model provides well localized atomic positions for GBs of various geometries.
\end{abstract}

\maketitle


\section{\label{intro} Introduction}

Grain boundaries are common defects in crystalline materials and play a major role in determining their physical, mechanical, electrical, and chemical properties. Also, modeling the segregation of solute atoms at grain boundaries (GBs) in steels is of great importance for the prediction of lifetimes of service materials. Several modeling approaches to this problem are available today [\onlinecite{Lejcek-book}]. Notably, a significant progress has been done using molecular dynamics (MD), where GB properties (segregation energies, grain boundary energies, etc.) have been studied at the atomic level [\onlinecite{Tschopp2012,TschoppMcDowell2007a,TschoppMcDowell2007b}]. Nevertheless, the study of segregation at GBs by MD remains prohibitively expensive. Principally, since MD is constrained to deal with the scale of nanoseconds, whereas physical phenomena related to diffusion of atoms take place at mesoscopic time scales.

Recently, the phase-field-crystal (PFC) model for binary alloys has been applied to this class of problems [\onlinecite{GreenwoodEtAl2011, StolleProvatas2013}] and has proven to capture the basic features of the GB segregation as well as the influence of undercooling, average alloy concentration, energy of mixing, etc. In spite of its big success in the description of GB properties, some limitations may stand in the way of the PFC method's applications to particular problems. Such problems could be the atomic density field smeared in topologically disordered regions, such as GBs, of the system. As a consequence, it can be impossible to localize all atoms forming a GB or even to determine their number.

In order to study either solute/impurity segregation or relaxation of GBs under a vacancy flux, the equilibrium atomic configurations of the GBs (with desired geometry) must be first obtained. Therefore, thousands of initial configurations should be tested in order to determine the lowest-energy ones [\onlinecite{Tschopp2012, TschoppMcDowell2007a, TschoppMcDowell2007b}]. In this paper, we show that it is now possible to avoid such a heavy and time-consuming work by using a new formulation of the atomic density function (ADF) model, introduced in Ref. [\onlinecite{Jin-Khachaturyan-2006}]. It will be shown that this method, when applied to modeling of GBs, gives atomic scale configurations very close to the equilibrium ones.

A physical interpretation of the atomic density function $\rho({\bf r})$ was done in Ref. [\onlinecite{Jin-Khachaturyan-2006}]. It has been shown that the ADF approach is a continuum limit of the old ADF theory based on the Onsager microscopic diffusion equation and in which atom positions are confined to the Ising lattice [\onlinecite{Khachaturyan_book}]. The small parameter in this case, is the ratio between the underlying Ising lattice parameter and the atomic interaction radius. Inspired by the mixing entropy of an alloy, the free energy is assumed here to be a functional of $\rho({\bf r})$ with a different entropic (local) term from [\onlinecite{Jin-Khachaturyan-2006}] and [\onlinecite{Elder-Grant-2004}]. This prevents the ADF from taking negative values since restricted to the interval $(0,1)$. From this point of view, our model presents the same virtues as the modified PFC model [\onlinecite{ChanGoldenfeldDantzig-2009, Berry-Grant-2011, Robbins-et-al-2012}], in which the atomic density function was restricted to positive values using an artificial penalizing term in the free energy. 

Finally, the nonlocal part of the free energy, which describes the interatomic interactions, is chosen to reproduce closely the first peak of the structure factor, as in Refs. [\onlinecite{Elder-et-al-2006, Wu-Karma-2006, Wu-Karma-2007, Jaatinen-et-al-2009, Jaatinen-et-al-2010}]. It has been shown, for example, that such a fit leads to a reasonable estimation of the elastic constants [\onlinecite{Jaatinen-et-al-2009}] and a correct anisotropy of the interfacial free energy [\onlinecite{Wu-Karma-2006}] for iron. The above allows us to present our results as specific for iron. Accordingly, we use the EAM (embedded-atom method) potential for iron [\onlinecite{Mendelev}] and the quench molecular dynamics in order to test the output of the ADF model.

Therefore, the atomic configurations obtained by means of the ADF model are subsequently relaxed in MD simulations and are shown to be quite close to the equilibrium configurations. It should be pointed out that the ADF model used in this study has the virtue of producing well localized atomic positions at all parts of the GB irrespectively of the geometry of the latter. The atomic positions thus can be directly imported into MD simulations. Although these configurations may prove to be somewhat compressed (when an EAM potential is applied), the structural units constituting the GBs are always reproduced correctly. In this paper, symmetrical [100] and [110] tilt GBs in bcc iron were modeled. Since the ADF model is a very efficient method in terms of computation time, it presents an excellent tool to obtain the atomic configurations of GBs of arbitrary geometry in a rapid way.

\section{The Model and its fit to the structure factor of iron}\label{1}

The foundation of the ADF model lies on the concept of the atomic density function. This function is interpreted as the density of point-like atoms averaged over the thermal vibration time scale and thus assuming a continuous form, [\onlinecite{TupperGrant2008}]. In this logic, the density function, at a given point, can be interpreted as the probability to find an atom in the infinitesimal vicinity of this point.

In the ADF model, the free-energy functional can be presented as a sum of
nonlocal and local terms and defined as ([\onlinecite{Jin-Khachaturyan-2006}])
\begin{eqnarray}\label{f_tot}
f(\{\rho({\bf r})\}) &=& \int \frac{d^3k}{(2\pi)^3}V(k)\rho_{\bf k}\rho_{-\bf k} + f_{\mathrm{loc.}},
\end{eqnarray}
where
\be\label{rho_k}
\rho_{\bf k} = \int \frac{d^3r}{\mathcal{V}} e^{i{\bf kr}}\rho({\bf r}),
\ee
is the Fourier transforms of the atomic density function ($\mathcal{V}$ is the volume of the system) and $V(k)=\int (d^3r/\mathcal{V})e^{i{\bf kr}}W(r)$ is the Fourier transforms of the interatomic potential $W(|{\bf r-r'}|)$. $f_{\mathrm{loc.}}$ is the local term of the free energy.

It is instructive to represent the first term in (\ref{f_tot}), using (\ref{rho_k}), as $\int d^3r \rho({\bf r})V(-i\nabla)\rho({\bf r})$ (where $\nabla$ should be understood as an operator acting on the function following it). Expanding $V(-i\nabla)$ up to the fourth order (in powers of the gradient operator) and omitting odd power terms (because of the central symmetry) one arrives at a form identical to the nonlocal term of the PFC model free energy functional [\onlinecite{Elder-Grant-2004}]. Using a nonexpanded form (\ref{f_tot}) is thus a generalization compared to the PFC model. 

Contrary to the initial version of the ADF theory [\onlinecite{Jin-Khachaturyan-2006}], as well as to the PFC model (where the local free energy was expressed in a polynomial form) in our model we express the local (entropic) term of the free-energy functional using a logarithmic form analogous to that of the binary alloy model:
\be\label{f_loc}
{\displaystyle f_{\mathrm{loc.}} = k_B T \int \frac{d^3r}{\mathcal{V}}\big[\rho({\bf r})\ln\rho({\bf r}) + (1-\rho({\bf r}))\ln(1-\rho({\bf r}))\big].}
\ee

In a crystalline phase the function $\rho({\bf r})$ presents a periodic set of peaks indicating equilibrium atomic positions. It should be realized, though, that the density of ``vacancies", $1-\rho({\bf r})$ in (\ref{f_loc}), is merely a probability to {\em not} find an atom at a point $\bf r$. It does not suppose a presence of localized vacancies in the atomic lattice formed by the peaks in $\rho({\bf r})$, which anyway would not have sense for a time-averaged picture.

When the system is unstable or metastable (with respect to a crystalline phase formation) and when the periodic fluctuations start to grow, the local part of the free energy restricts the amplitude of the peaks near $0$ and $1$. This results in distinct atomic density peaks of equal height (nearly $1$) even for the ``atoms" in such unfavorable positions as grain boundaries. The liquid phase corresponds to a uniform $\rho({\bf r})=\text{const}$, i.e. $\rho({\bf r})=\rho_{{\bf k}=0}$ [see (\ref{rho_k})]. To investigate the stability of a liquid state with respect to density modulations, we should expand the free energy functional, Eq. (\ref{f_tot}), into the Taylor series with respect to the spatial variations of the atomic density $\Delta\rho({\bf r}) = \rho({\bf r}) - \rho_0$. In analogy with the theory of crystal lattice vibrations, the quadratic term of this expansion will be called the harmonic term. The harmonic term can be written in terms of the density wave amplitudes $\Delta\rho_{\bf k}$ as:
\begin{eqnarray}\label{Delta_f}
\Delta f &=& \int \frac{d^3k}{(2\pi)^3}D(k)\Delta\rho_{\bf k}\Delta\rho_{-\bf k},
\end{eqnarray}
where the response function $D(k)$ is the second functional derivative of the free energy functional (\ref{f_tot}):
\be\label{D(k)}
D(k) = \delta^2 f(\{\rho_{\bf k}\})/{\delta\rho_{\bf k}}^2.
\ee
Its minimal value, $D(k_0)=\min(D(k))$, describes stability of a given liquid state characterized by a uniform ADF with $\rho({\bf r})=\rho_0$. A negative value of $D(k_0)$ corresponds to a liquid state absolutely unstable with respect to solid state formation: any infinitesimal fluctuation of a uniform ADF would lead to a periodic structure growth. The lattice spacing of the latter will be related to the minimum position $k_0$ as $a=2\sqrt{2}\pi/k_0$ (for a bcc structure). Positive $D(k_0)$ corresponds to a globally or locally stable liquid phase.

As it was pointed out in Ref. [\onlinecite{Jin-Khachaturyan-2006}] as well as in Ref. [\onlinecite{Elder-Grant-2004}], the response function $D(k)$ can be related directly to the structure factor according to:
\be\label{S(k)->D(k)}
S(k) = k_B T D^{-1}(k).
\ee
From (\ref{D(k)}) one finds $D(k)=V(k)+k_BT/(\rho_0(1-\rho_0))$. It is convenient to choose $V(k)$ in the form [\onlinecite{Jin-Khachaturyan-2006}]
\be\label{V(k)}
V(k) = V_0\left(1-k^4/\left[(k^2-{k_1}^2)^2+{k_2}^4\right]\right),
\ee
so that the values of the parameters $V_0$, $k_1$, $k_2$, and the average of ADF ($\rho_0$), can be fitted knowing the structure factor of the given material. The value $k=k_0$ minimizing $V(k)$ is given by $k_0=\sqrt{{k_1}^4+{k_2}^4}/k_1$.

\begin{figure}
\center{\includegraphics[width=0.3\textwidth,angle=-90]{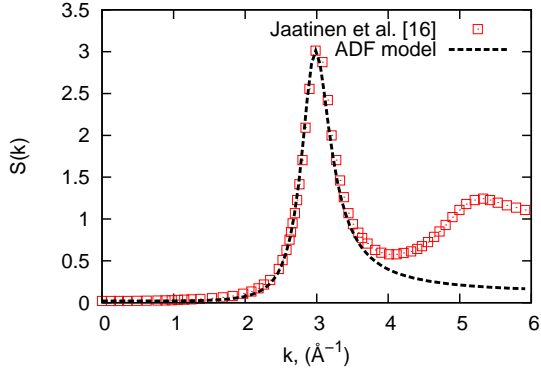}}
\caption{Fit of the structure factor of iron at the melting point in the ADF model (dashed line). The squares represent the data obtained by conversion of the correlation function $C(k)$ (used in Ref. [\onlinecite{Jaatinen-et-al-2009}])  to the structure factor according to the relation $S(k)=[1-C(k)]^{-1}$ (Ref. [\onlinecite{Chaikin-Lubensky}]).}\label{S(k)_fig}
\end{figure}

We have chosen to fix the parameters of $V(k)$ and the average value of the ADF by fitting the structure factor of our model to that of iron at its melting point. To do this, we used the function $S(k)$ following from the combination of MD simulations and experimental data as presented and explained in Ref. [\onlinecite{Jaatinen-et-al-2009}]. As in Ref. [\onlinecite{Jaatinen-et-al-2009}], we fitted the position $k_0 = 2.985 \AA^{-1}$ and the height $S(k_0)= 3.012$ of the maximum of $S(k)$, as well as its width $S''(k_0)=-94.35 \AA^2$ and the value $S(k=0)=0.02$ related to the isothermal compressibility of the liquid.

The formulas that relate the parameters of our model to $k_0$, $S(k_0)$, $S''(k_0)$, and $S(0)$ are the following:
\begin{eqnarray}
k_1 &=& \frac{k_0}{\sqrt{1-\frac{8}{{k_0}^2} \frac{S(k_0)}{S''(k_0)}\left(\frac{S(k_0)}{S(0)}-1\right)}},\label{k1}
\\
k_2 &=& k_0\frac{\sqrt[4]{-\frac{8}{{k_0}^2} \frac{S(k_0)}{S''(k_0)}\left(\frac{S(k_0)}{S(0)}-1\right)}}{\sqrt{1-\frac{8}{{k_0}^2} \frac{S(k_0)}{S''(k_0)}\left(\frac{S(k_0)}{S(0)}-1\right)}},\label{k2}
\\
\frac{V_0}{k_BT} &=& \frac{\left(\frac{S(k_0)}{S(0)}-1\right)^2}{S(k_0)\left(\frac{S(k_0)}{S(0)}-1\right)-S''(k_0){k_0}^2/8},\label{V0}
\\
\frac{1}{k_BT}\frac{\delta^2 f_{\mathrm{loc.}}}{\delta\rho^2} &=& \frac{\frac{S(k_0)}{S(0)}-1-S''(k_0){k_0}^2/8}{S(k_0)\left(\frac{S(k_0)}{S(0)}-1\right)-S''(k_0){k_0}^2/8}.\label{d2f/dr2}
\end{eqnarray}
Note that, since $S(0)>0$, $S(k_0)>S(0)$, and $S''(k_0)<0$, all parameters defined by Eqs.(\ref{k1})--(\ref{d2f/dr2}) take real positive values. Note that, if $f_{\mathrm{loc.}}$ is given by (\ref{f_loc}), $\delta^2 f_{\mathrm{loc.}}/\delta\rho^2=k_BT/(\rho_0(1-\rho_0))$, where $\rho_0$ is the average value of the atomic density function. The fit of the structure factor in Fig. \ref{S(k)_fig} leads to $k_1=0.434868 k_0$, $k_2=0.625776k_0$, $k_BT/V_0=0.024829$, $\rho_0=0.116372$.

\section{Symmetrical tilt GB\MakeLowercase{s} in ADF model and MD simulations}

Grain boundaries of desired geometry are obtained by crystallizing a liquid layer placed in-between two crystal grains of chosen orientation (in fact, due to the periodic boundary conditions, two identical GBs are formed in this way). The temperature and the mean value of the atomic density function were chosen for simplicity as $k_BT=0.025V_0$ and $\rho_0=0.1$, which are approximately the same values as those fitting the structure factor of iron. The parameters of the atomic potential were taken exactly as those fitted in the previous section. One can check that $D(k_0)(\simeq0.0445)$ is positive, so that the liquid at this temperature and density is metastable with respect to infinitesimal fluctuations of the ADF. A stable crystalline phase can be formed in the system as long as a sufficiently large crystalline nucleus is taken as the initial state.

The model is discretized in order to be treated numerically and the resolution of the numerical grid is chosen so that the lattice constant would contain 16 grid cells: $a=16\Delta x$ ($\Delta x=1$).

The GB orientation with respect to the simulation box is chosen such that the normal to the GB plane is along the $Oz$ axis of the box, and the tilt axis is along $Ox$. The choice of different dimensions of the simulation box is then determined by different factors. Since the GB does not affect in any way the periodicity of the bicrystal in the $Ox$ direction (the period being $a$ and $\sqrt{2}a$ for $[100]$ and $[110]$ tilt GBs, respectively), this dimension can be taken as small as a few atomic planes, in order to save computational resources. The dimension $Oy$ is determined by the pattern of the GB and therefore is varied according to the tilt angle. Finally, as one deals with a pair of identical GBs due to periodic boundary conditions, the $Oz$ dimension is determined by the distance one wants to have between the GBs. While it seems to be preferable to keep this distance as large as possible to have a better estimation of an isolated GB energy, the argument presented in the following suggests to tend to rather moderate dimensions $Oz$.

The total free energy (\ref{f_tot})-(\ref{f_loc}) can be decomposed into
\be\label{f_tot_2}
f_{\mathrm{tot}} = f_{\mathrm{bulk}} + f_{\mathrm{GB}}S_{\mathrm{GB}}/\mathcal{V},
\ee
where $f_{\mathrm{bulk}}$ is the bulk energy per unit volume, $f_{\mathrm{GB}}$ is the surface energy of the totality of GBs of surface $S_{\mathrm{GB}}$, present in the system, and $\mathcal{V}$ is the total volume of the system. Since the ratio $S_{\mathrm{GB}}/\mathcal{V}$ has quite a small value, the estimation of $f_{\mathrm{GB}}$ basing on $f_{\mathrm{tot}}$ should be done with big caution. This forced us to stick with a rather small $Oz$ dimension, sometimes not exceeding $\simeq$ $5$ nm, which makes only $2.5$ nm between the neighboring GBs. Nevertheless, we have verified that such a small distance does not change either the structure or the energy of the GB in any significant way, i.e., gives results representative of larger grain sizes. In principle, it is sufficient to know the bulk energy per volume and the geometry of the GB to obtain the $f_{\mathrm{GB}}$ from (\ref{f_tot_2}). In reality, however, it turns out that the variations in $f_{\mathrm{bulk}}$ due to the discreteness of the simulation grid risk to cloud the second term in (\ref{f_tot_2}). It is to avoid this source of error, that we chose the relatively high value of 16 grid units per one BCC lattice spacing $a$.

The modern theory of high-angle grain boundaries is based on the structural unit model, first introduced in Ref. [\onlinecite{SuttonVitek1983}] and confirmed since using computer experiments [\onlinecite{WangSuttonVitek1984, Tschopp2012}] and also high-resolution transmission electron microscopy (HRTEM) images (for example, Ref. [\onlinecite{MerkleSmith1987}]). This model allows one to explain the peculiarities of the GB energy behavior at some misorientation angles [\onlinecite{Gui-JinVitek1986}]. According to it, each GB characterized by a certain sigma number (the inverse of the number of coincident lattice sites in the two grains, when the grains are superposed) can be decomposed in a periodic repetition of a pattern consisting of elementary structural units. Some structural units have energies lower than the others; a number of high-angle GBs, composed uniquely by these low-energy units, have energies notably lower than the other GBs in the same tilt angle range, and hence are referred to as special GBs. GBs close to the special ones are a mixture of low- and high-energy structural units. It was demonstrated in Ref. [\onlinecite{Gui-JinVitek1986}] that the energy in the vicinity of special GBs follows a logarithmic law analogous to that derived by Read and Shockley for low-angle GBs [\onlinecite{ReadShockley}].

Following, we will present the results of ADF modeling of [100] and [110] tilt GBs of different angles. The main quantitative characteristic that we looked for is the GB energy. Another important way to characterize GBs obtained by the ADF method is to identify typical patterns of structural units that are quite well known today from MD simulations. Moreover, the GB configurations we had obtained were subsequently relaxed using the Fe EAM potential developed by Mendelev {\it et al.} [\onlinecite{Mendelev}]. (Note that it is the same potential that was used to obtain the structure factor in Fig. \ref{S(k)_fig} [\onlinecite{Jaatinen-et-al-2009}].) The initial atomic position of each atom was discretized on a grid; the simplicity of this procedure is due to atomic density peaks with well pronounced centers (where the ADF approaches 1 most closely). The number of atoms was also provided from the ADF configurations. The quench MD method was used to relax (at 0 K temperature) each configuration at constant volume with periodic boundary conditions. In a second step, the simulation box was relaxed only in the direction normal to the GB habit plane (the simulation box is fixed according to the tilt angle and the equilibrium lattice parameter in the two other directions) in order to minimize the energy of the system.

\begin{figure}
\center{\includegraphics[angle=-90,width=0.5\textwidth]{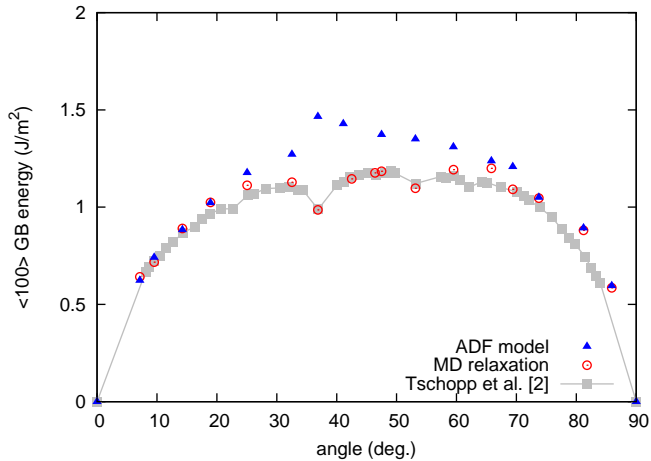}}
\caption{The energy of $[100]$ tilt GBs. The energy scale in the ADF model is chosen to fit the low-angle GBs' energies after MD relaxation.}
\label{energy_100}
\end{figure}

\begin{figure}
\center{\includegraphics[angle=0,width=0.5\textwidth]{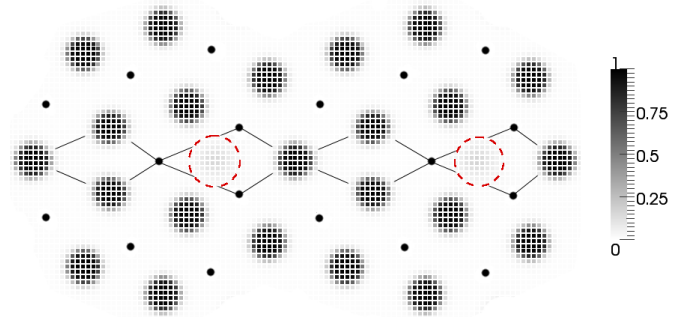}}
\caption{A cross section of the atomic density profile of a $\Sigma 5 (031)$ $(36.87^\circ)$ $[100]$ tilt GB in the plane $(100)$. The dots represent atomic positions in the adjacent plane, in order to visualize the structural units. The ``bumps'' that appear in the empty space are highlighted.}
\label{Sigma_5}
\end{figure}

We will first consider [100] tilt GBs. In Fig. \ref{energy_100}, the energy of [100] tilt GBs is presented for different tilt angles. Let us remark that grain boundaries with the tilt axis $[100]$ do not possess such remarkable energy cusps as those present in $[110]$ tilt GBs [\onlinecite{Tschopp2012}]. Since the energy scale has not been chosen yet in our model, this is done now, in order to compare the ADF and MD results, by fitting the energy of low-angle GBs. A remarkable coincidence between the ADF data and the relaxed energies in the mentioned range becomes evident once the energy scale has been chosen properly. There is however some notable energy overestimation by the ADF model in the middle range of angles, especially in the vicinity of the angle $36.87^\circ$ ($\Sigma 5 (031)$). This angle corresponds to the most remarkable energy cusp of [100] tilt GBs, where a minimum {\em should} have been obtained instead of the maximum we observed. We suppose, nevertheless, that the atomic configurations are correct (since they reveal correct structural units and lead to correct energies after MD relaxation). We explain this disagreement by a contribution of ``bumps'' which can be noticed (Fig. \ref{Sigma_5}) in poorly packed regions of the GB. Their height is about $10\%$ of that of atomic peaks. Those low bumps are therefore not interpreted as atoms when importing the atomic positions to MD simulation, otherwise the system would be over-compressed. Such bumps are noticeable also for other high-angle [100] tilt GBs, but are most concentrated in the $\Sigma 5 (031)$ GB. We have observed a correlation between the ``frequency'' of low bumps along the GB and the overestimation of the energy by the ADF method.

Let us remind the reader that in the ADF model the atomic density function represents the probability to find an atom at a given site ${\bf r}$ and bumps can be interpreted as sites where this probability is much smaller but not negligible with respect to the ``normal'' atomic position. Since we perform the simulation at a high temperature, it is not surprising that the atoms have a tendency to go on a vacant site and spend some time in this position. As a consequence of the excess of poorly packed regions in $\Sigma 5 (031)$ GB we were able to obtain a structural unit pattern alternative to that presented in Fig. \ref{Sigma_5} by expanding/contracting the GB perpendicularly to the GB plane. This second configuration was, however, rejected for giving a too high energy in MD simulations.

This problem does not arise for [110] tilt GBs, since (110) atomic planes are significantly more dense packed comparing to (100) planes (by a factor of $\sqrt{2}$). [110] tilt covers a range from $0^\circ$ to $180^\circ$. The ADF model led to unique structural unit patterns for the reference angles that we have compared to [\onlinecite{Tschopp2012}]. Rigid body translation perpendicular to the GB plane (expansion/contraction) does not compromise the form of the structural units in this case. We have studied the influence of rigid body translation within the GB plane for a few different orientations for $[100]$ and $[110]$ tilts, and no new low-energy configurations were found. This will be confirmed further in the text by comparing our energies to those from [\onlinecite{Tschopp2012}] where in-plane translations {\em were} tested for all angles.

\begin{figure}
(a)\includegraphics[angle=0,width=0.15\textwidth]{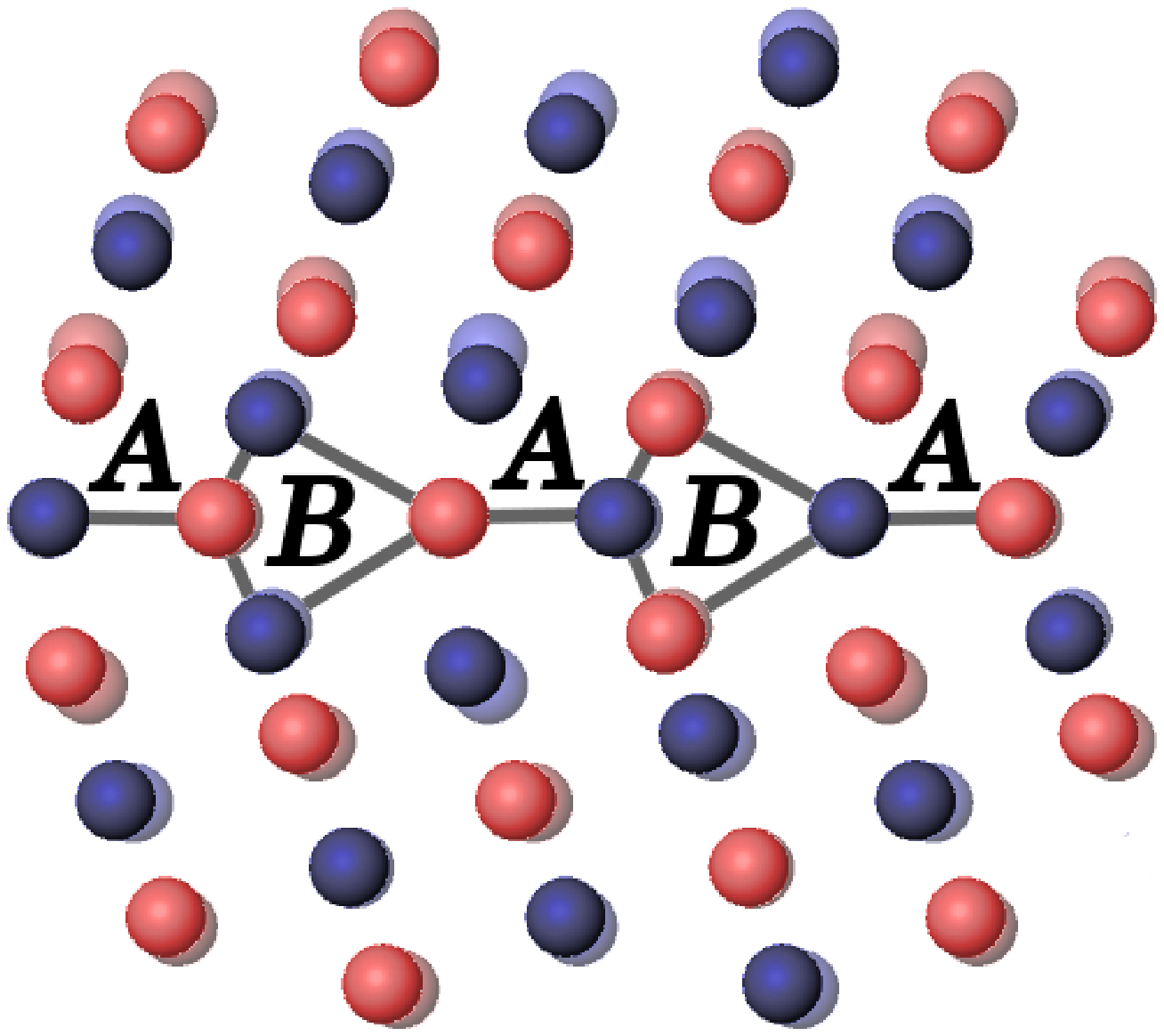}
(b)\includegraphics[angle=0,width=0.16\textwidth]{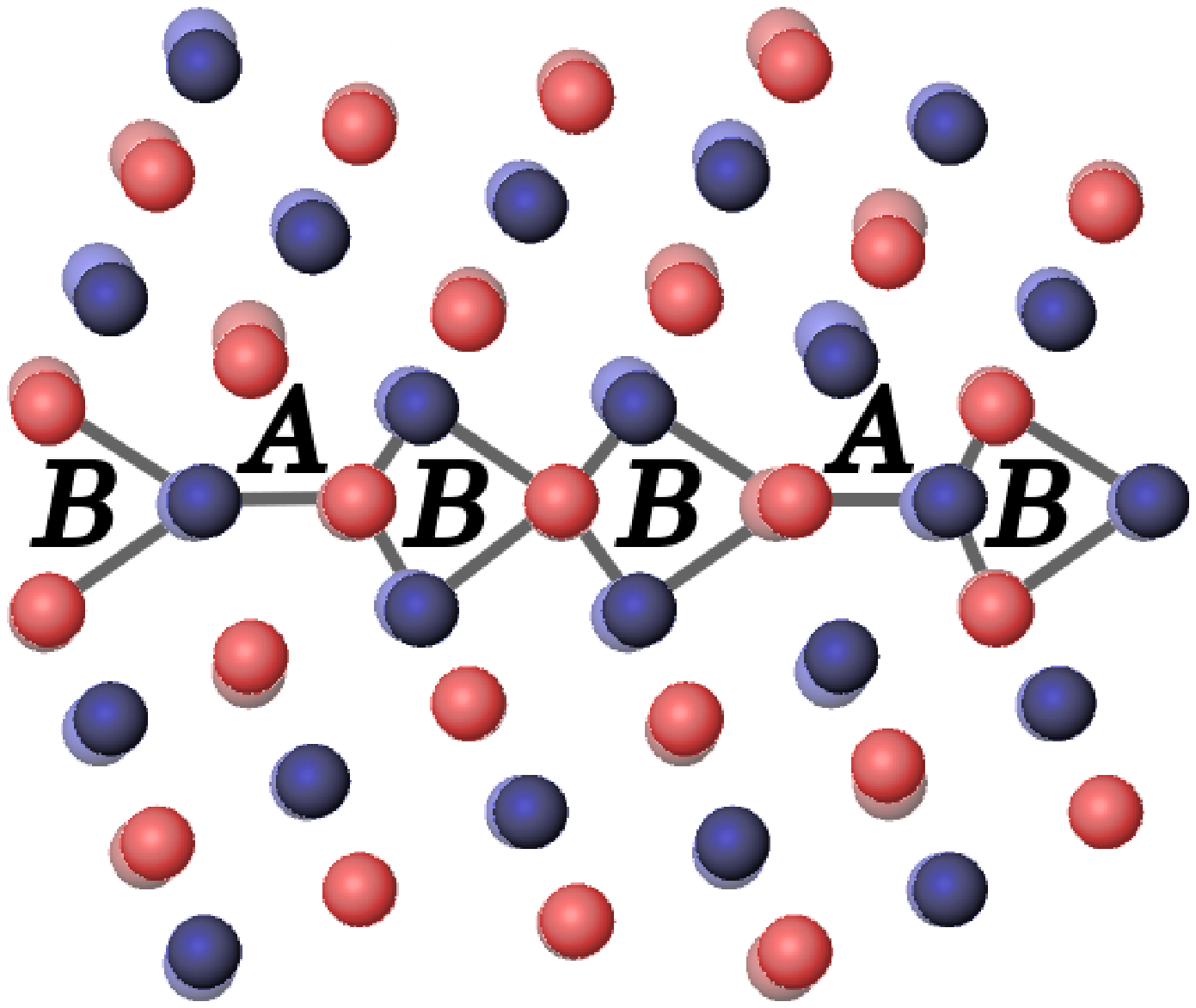}

\vspace{0.3cm}

(c)\includegraphics[angle=0,width=0.10\textwidth]{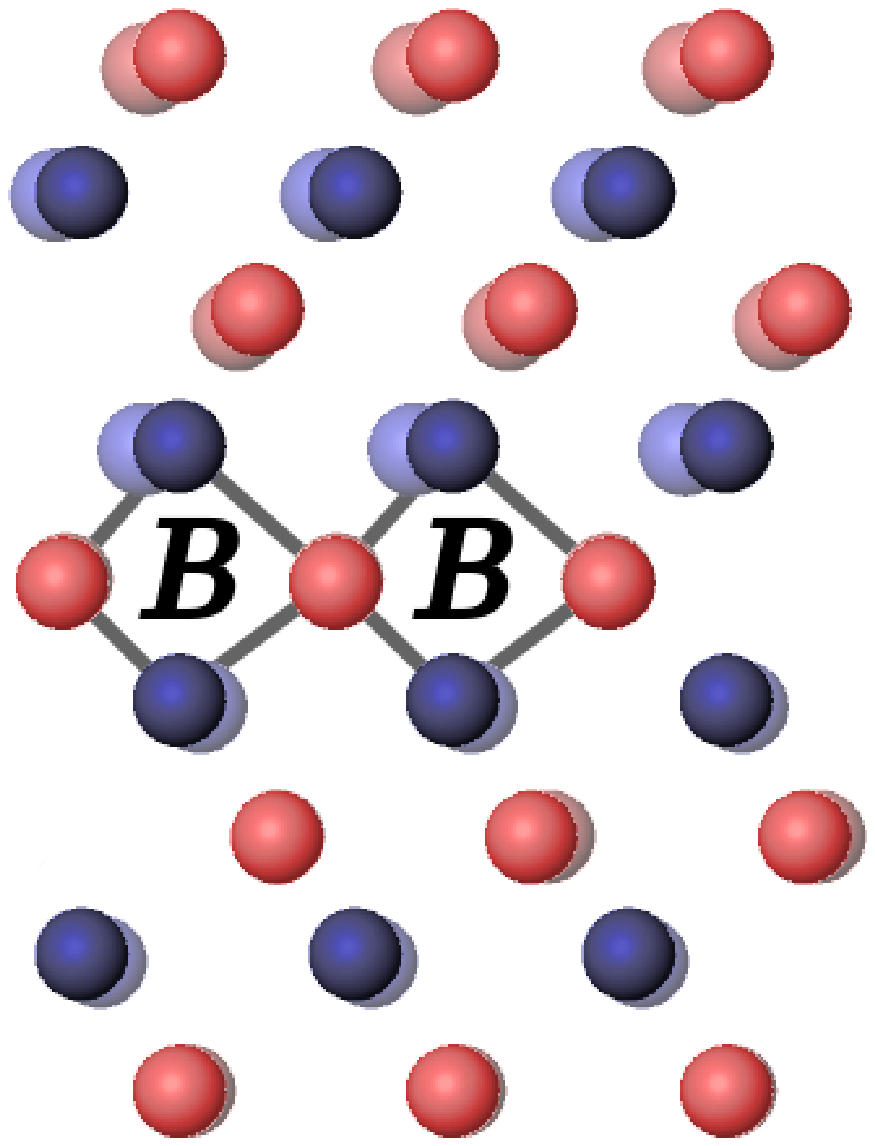}
(d)\includegraphics[angle=0,width=0.15\textwidth]{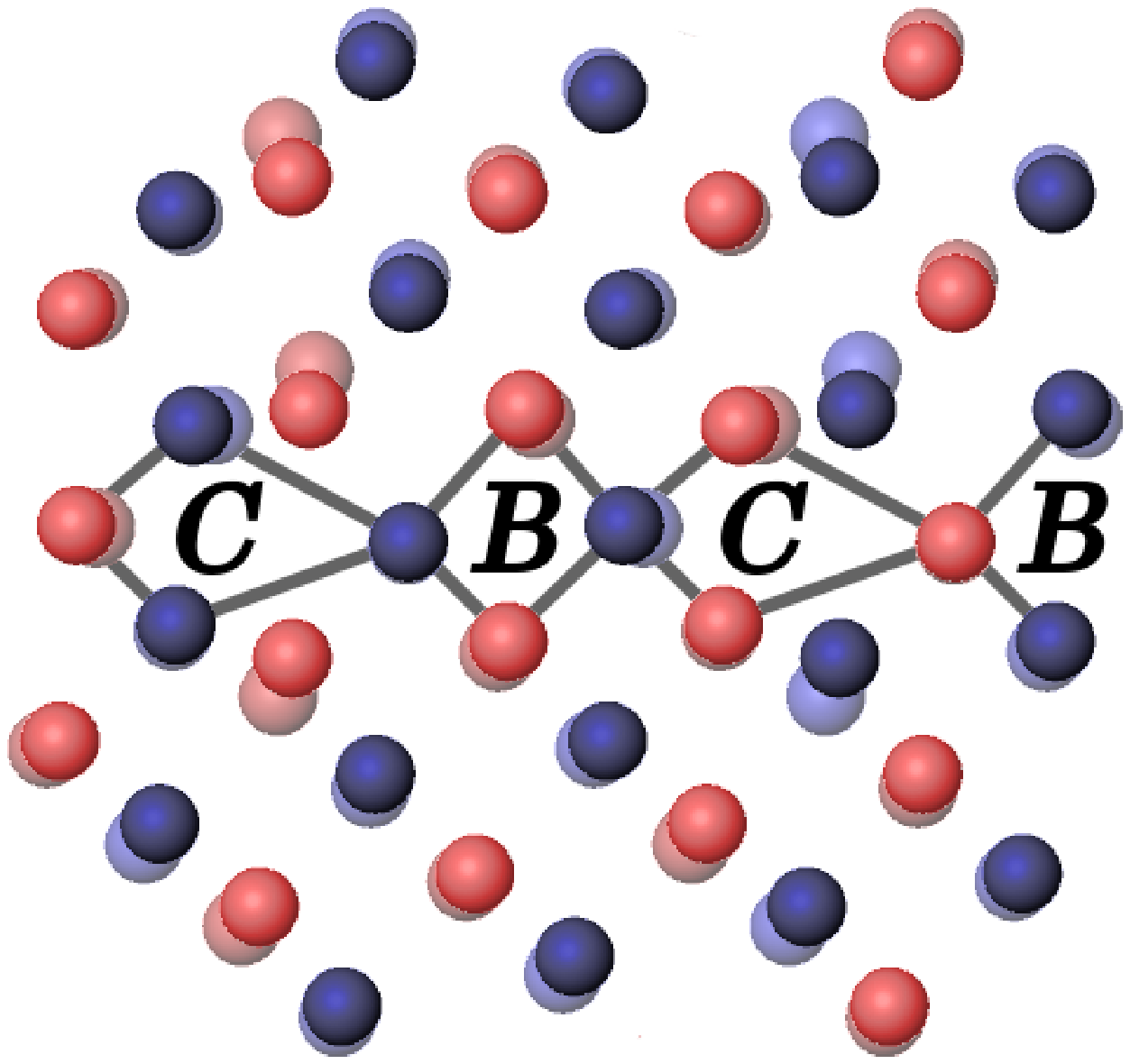}
(e)\includegraphics[angle=0,width=0.10\textwidth]{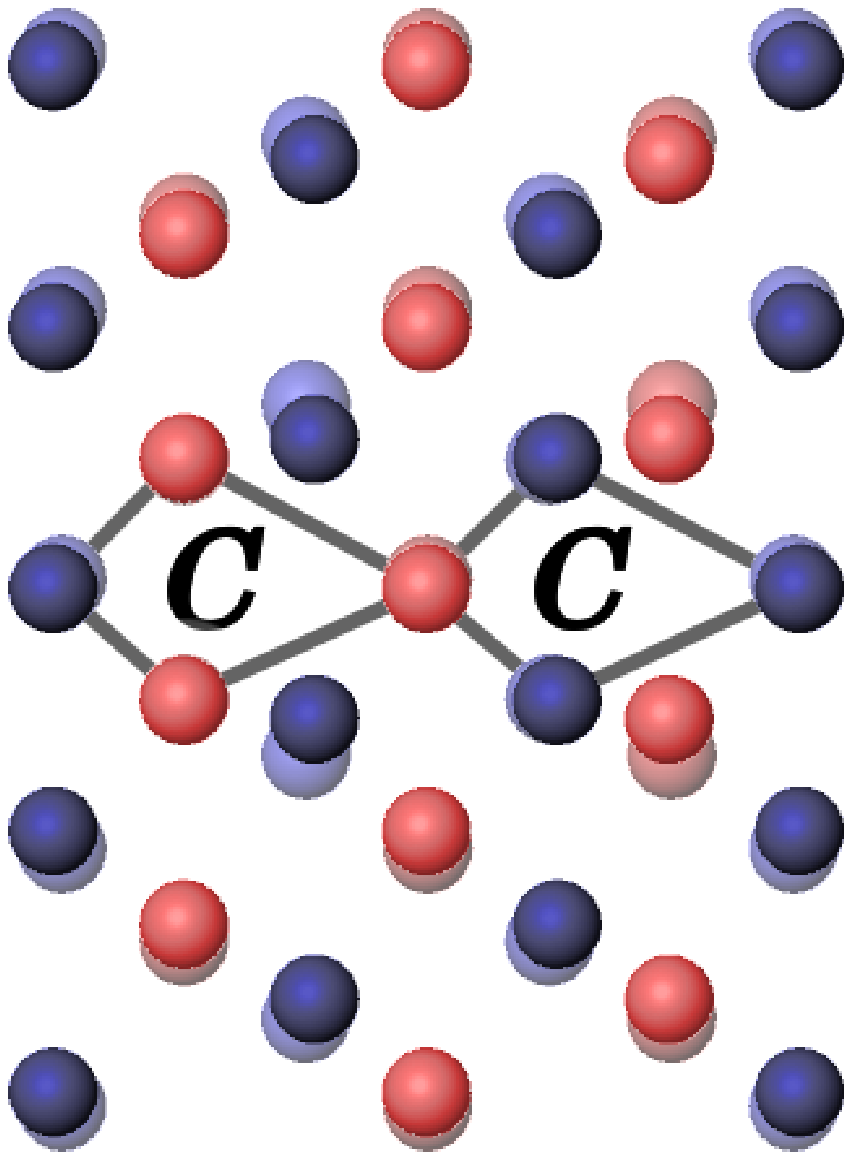}

\vspace{0.3cm}

(f)\includegraphics[angle=0,width=0.18\textwidth]{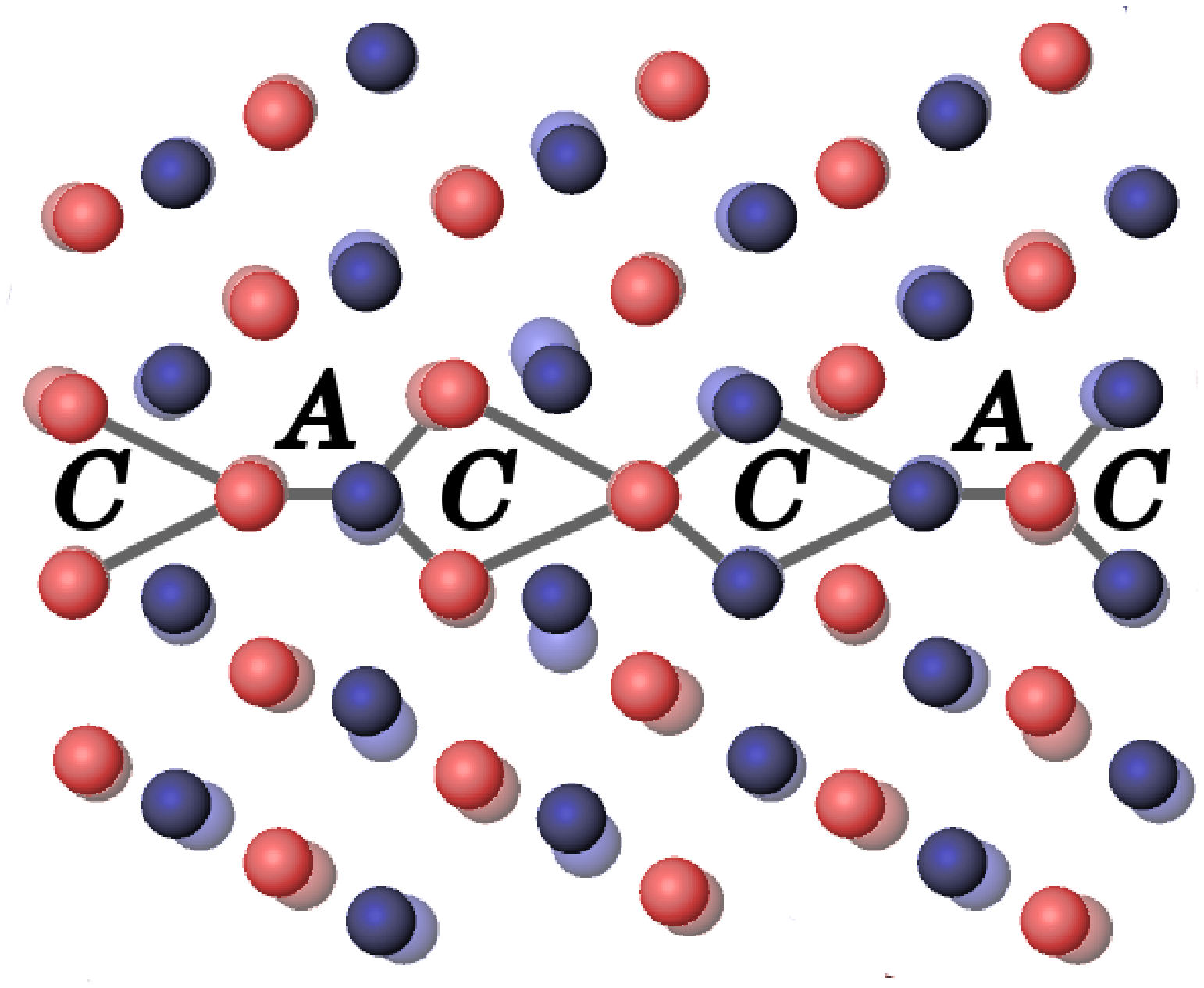}
(g)\includegraphics[angle=0,width=0.13\textwidth]{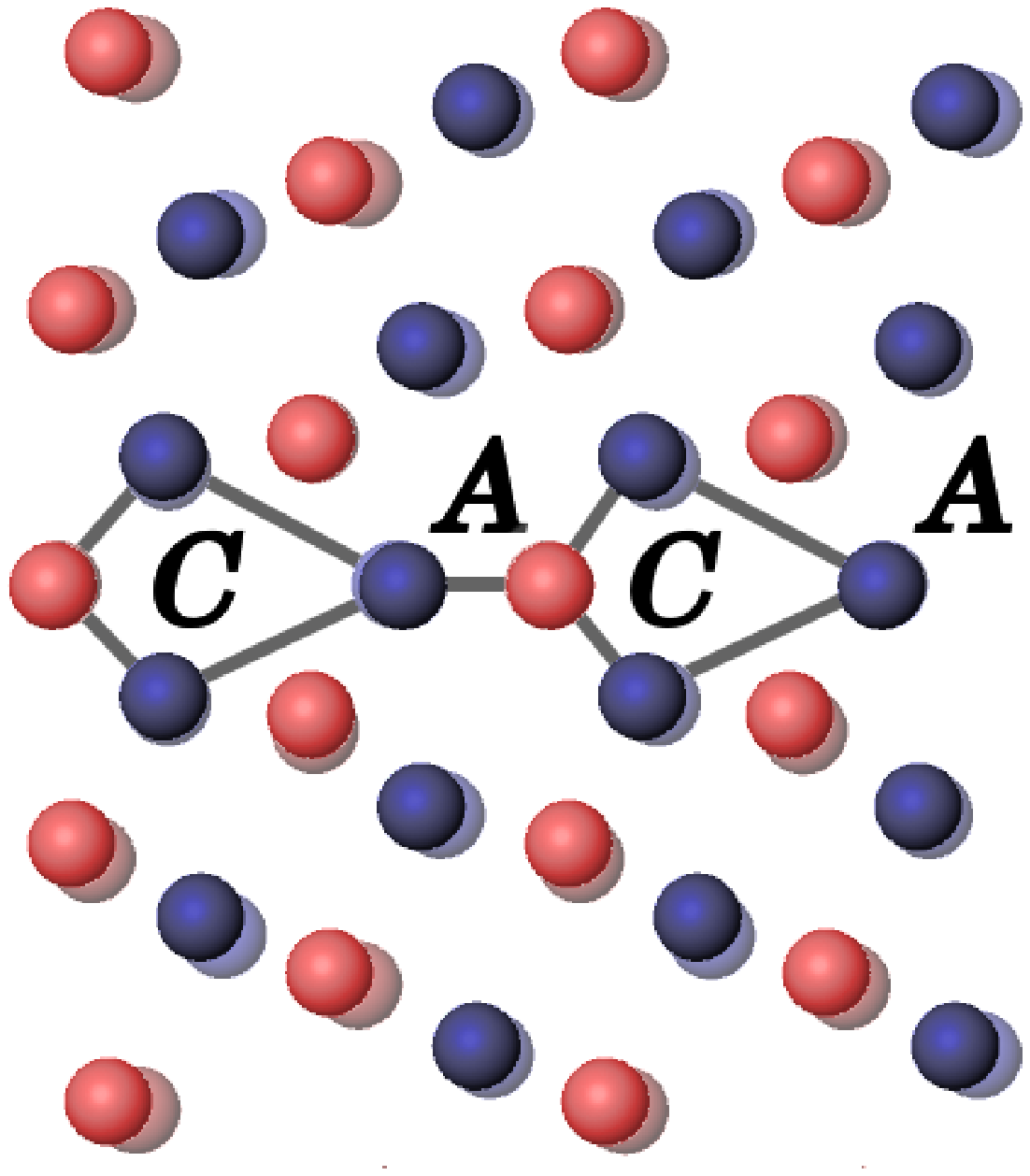}

(h)\includegraphics[angle=0,width=0.16\textwidth]{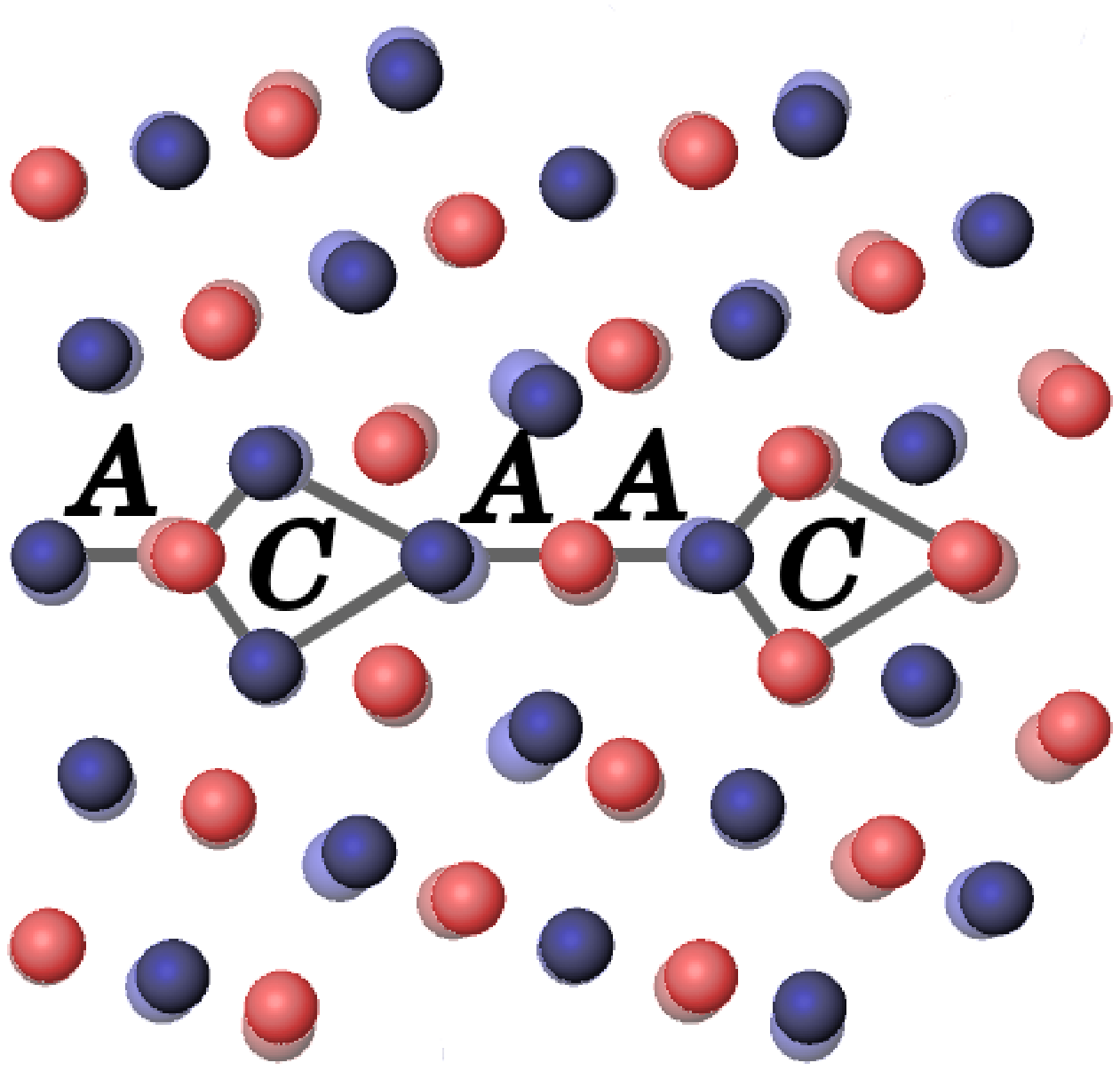}
(i)\includegraphics[angle=0,width=0.18\textwidth]{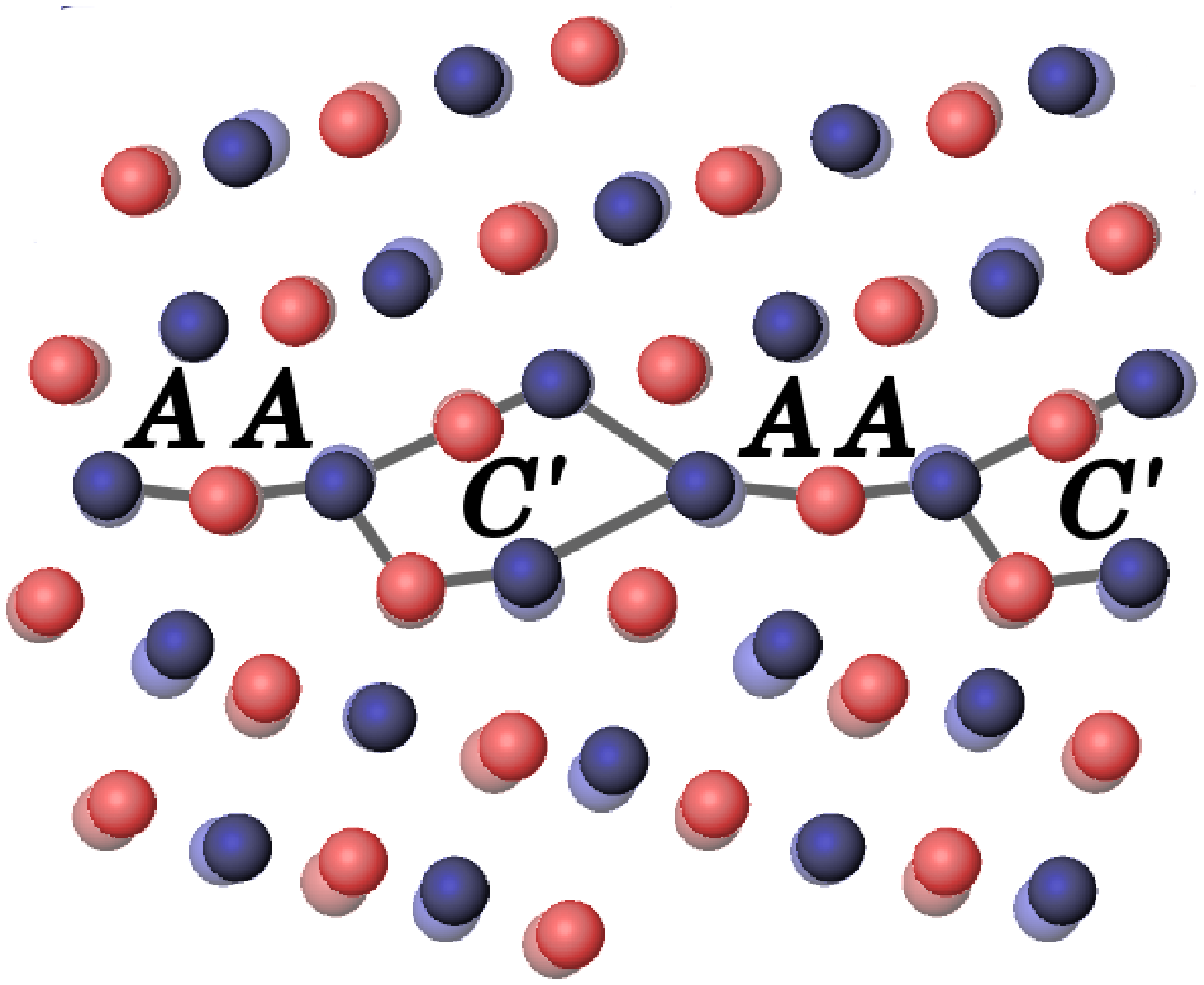}

\includegraphics[angle=0,width=0.22\textwidth]{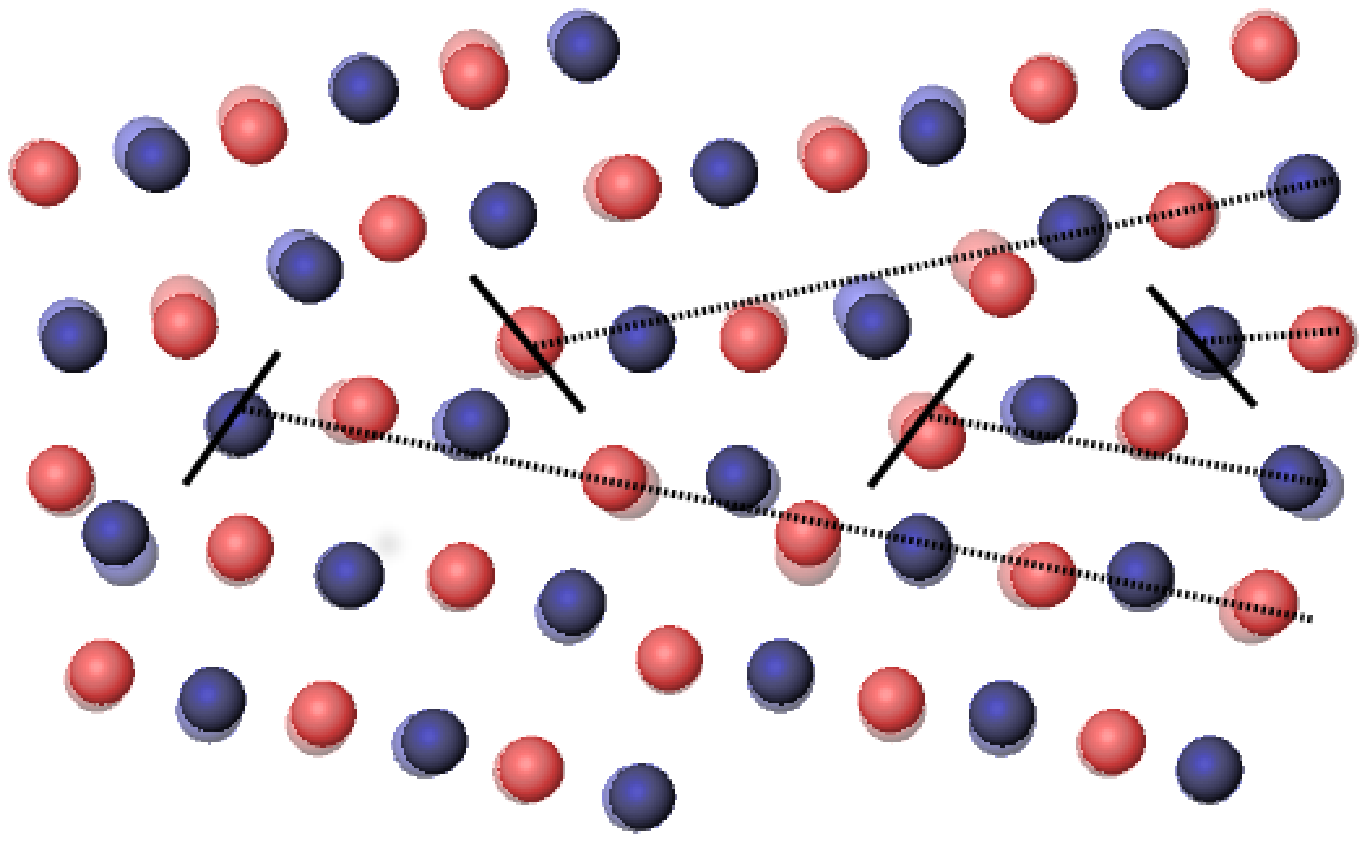}
\includegraphics[angle=0,width=0.22\textwidth]{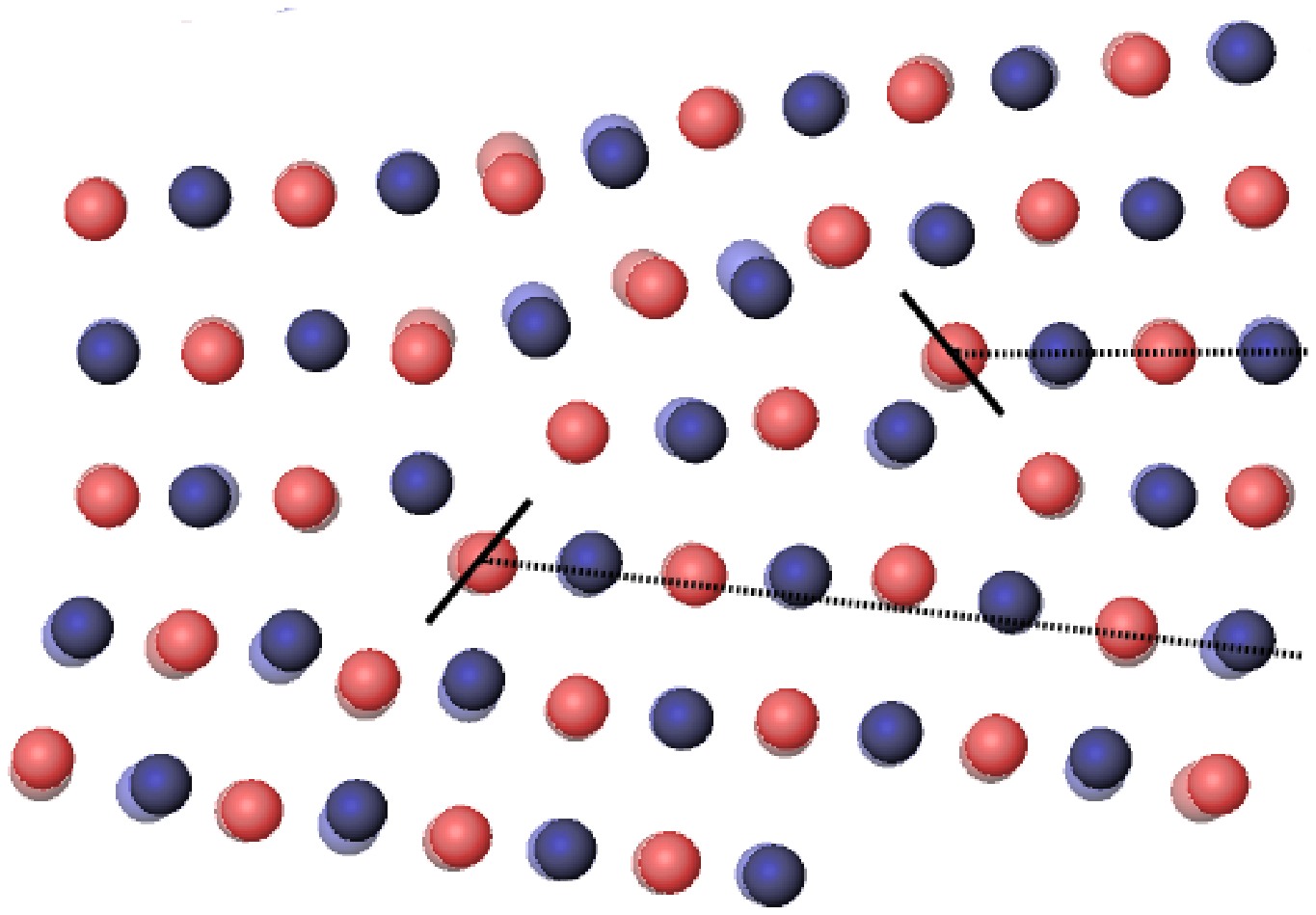}

(j)  \hspace{2.5cm} (k)

\caption{Structural units in $[110]$ tilt GBs modeled by the ADF: (a) $38.94^\circ, \Sigma 9 (114)$; (b) $50.48^\circ, \Sigma 11 (113)$; (c) $70.53^\circ, \Sigma 3 (112)$; (d) $93.37^\circ, \Sigma 17 (334)$; (e) $109.53^\circ, \Sigma 3 (111)$; (f) $121.01^\circ, \Sigma 33 (554)$; (g) $129.52^\circ, \Sigma 11 (332)$; (h) $141.06^\circ, \Sigma 9 (221)$; (i) $148.41^\circ, \Sigma 27 (552)$; (j) $153.47^\circ, \Sigma 19 (331)$; (k) $163.9^\circ, \Sigma 51 (551)$. Equilibrium atomic positions after relaxation in MD simulation are indicated in the background with faded colors; two colors refer to two adjacent atomic planes. In the last row, dislocations are highlighted.}\label{structural_units}
\end{figure}

The most direct way of testing the GB atomic structures we obtained is to compare our configurations to the well-established structural unit patterns for $[110]$ tilt GBs [\onlinecite{Tschopp2012}]. In Fig. \ref{structural_units}, structural unit composition of some $[110]$ tilt GBs is presented. The red and the blue atoms refer to two adjacent atomic planes. The atomic positions relaxed in MD simulations are indicated with faded colors in the background. Structural units are indicated with gray lines and their letter notations. The reader is referred to  Ref. [\onlinecite{Tschopp2012}] for atomic configurations of the same GBs obtained by MD simulation based on sampling of several thousands of initial states.

It is worth noting that we have obtained well-localized (and quite close to equilibrium) atomic positions for all the atoms forming GBs, even for the angles approaching $180^\circ$, where the regular PFC model seems to show a smeared density distribution on the GBs [\onlinecite{Jaatinen-et-al-2010}]. One can compare directly the angle $153.47^\circ [\Sigma 19 (331)]$ obtained with the two methods: Fig. \ref{structural_units}(j) and Ref. [\onlinecite{Jaatinen-et-al-2010}]. For the angles $153.47^\circ$ and $163.9^\circ$ [Figs. \ref{structural_units}(j) and \ref{structural_units}(k)], one can clearly see pile-ups of edge dislocations with Burgers vectors oriented at $54.74^\circ$ ($=\tan^{-1}\sqrt{2}$) with respect to the GB plane. From what it seems, the latter was not reproduced in the regular PFC model [\onlinecite{Jaatinen-et-al-2010}]. For low angles ($\leq 15^\circ$), one observes pile-ups of edge dislocations with Burgers vectors perpendicularly oriented to the GB plane; they are not shown here as this type of low-angle GBs is very widely presented in literature.

It is seen from Fig. \ref{structural_units}, that the most notable relaxation, when such is observed, happens on atoms adjacent to those composing the structural units and even sometimes those located deeper in the bulk. This is due to the fact that the configurations obtained by ADF modeling are more or less under compression, since, in the ADF model, atoms can change their ``size'' (the width of the atomic density function peaks) and approach each other more closely in the vicinity of the GB. This was confirmed by MD relaxation of the ADF's simulation box in the $Oz$ direction (perpendicularly to the GB). The equilibrium atomic volume was found to exceed the atomic volume obtained in the ADF modeling.

\begin{figure}
\center{\includegraphics[angle=-90,width=0.5\textwidth]{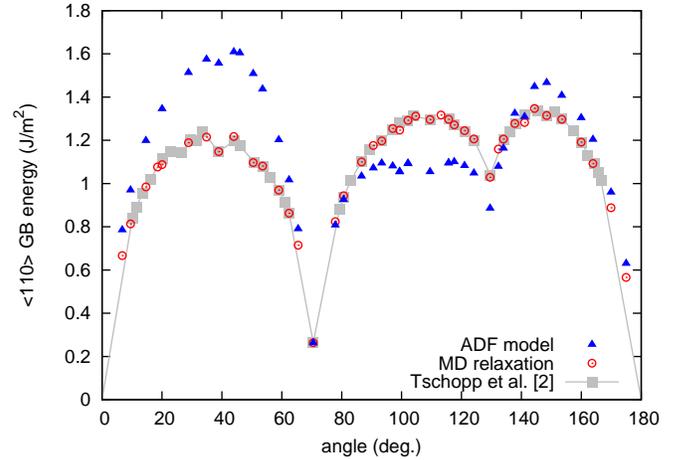}}
\caption{The energy of $[110]$ tilt GBs. The energy scale for the ADF data is chosen such that the $\Sigma 3 (112)$ energy coincides with its MD value.}
\label{energy_110}
\end{figure}

When ADF configurations are relaxed in MD keeping the same dimensions of the simulation box, the mentioned compression, distributed in the entire volume, does not affect the GB energy drastically. After relaxation of the simulation box, the energy curve practically coincides with that obtained in Ref. [\onlinecite{Tschopp2012}] (see Fig. \ref{energy_110}). Some fairly minor deviations for a few angles are probably due to slight tensions in the simulation box resulting from discretization error.

The energy scale for the ADF data presented in Fig. \ref{energy_110} was chosen in order to fit the $\Sigma 3 (112)$ ($70.53^\circ$) $[110]$ tilt GB energy. Reasons for this choice will be given in detail below. We have preferred, however, to keep in Fig. \ref{energy_100} the scale which made obvious the identical energy tendencies of low-angle $[100]$ tilt GBs in the ADF and MD methods (the two scales relating as $E_{110}/E_{100}\simeq 1.288$).

A possible source of discrepancy between the ADF and MD modeling is the difference in temperature. The ADF modeling was done at a temperature that is close to that of the iron melting point, while the MD relaxation was done at 0 K. In order to superpose the energy scales in these two simulations, we had to fit the energy of the GB for which the energy variation with temperature is expected to be minimal. It was shown in Ref. [\onlinecite{Biscondi}] that it is the case for the $\Sigma 3 (112)$ $[110]$ GB. Therefore, the fit of the energy scale was done based on the energy of this particular GB. Its value was previously rectified by taking the exact equilibrium dimension of the simulation box from MD. This choice was also dictated by the fact that this tilt angle results in a configuration very close to that of a regular crystal and thus has more reasons to be independent of the model applied.

\begin{figure}
\center{\includegraphics[angle=-90,width=0.45\textwidth]{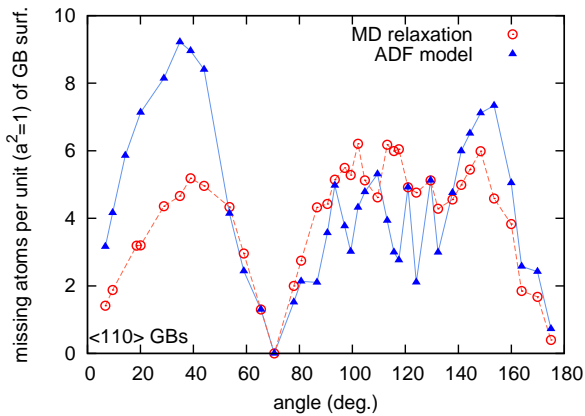}}
\caption{The average number of atoms missing from the first two coordination spheres of $[110]$ tilt GBs (the interval taken roughly as $0.81a<r<1.2a$), calculated per unit area ($a$ taken as the unity of length). Note that fluctuations in the ADF data can be explained in part by an error in atom position determination on a discrete simulation grid.}
\label{missing_atoms_110}
\end{figure}

The latter statement can be justified from Fig. \ref{missing_atoms_110} where the average number of atoms (per unit GB surface) missing from the first and second coordination spheres near the GB is presented as a function of the tilt angle. The first coordination sphere is located at $r=0.86a$, the second one at $r=a$. We have considered, roughly, as contributing to the first two coordination spheres, interatomic distances $r$ in the range $0.81a<r<1.2a$ ($1.2a$ is in the middle between the second and the third coordination spheres). As can be seen from Fig. \ref{missing_atoms_110}, the two curves obtained from MD and ADF modeling configurations coincide for the $\Sigma 3 (112)$ GB ($70.53^\circ$). Therefore, Fig. \ref{missing_atoms_110} suggests that the two methods have a less pronounced discrepancy for this particular angle, at least, as far as the energy is concerned. Indeed, the latter is determined by the relative positions of atoms. Choosing the energy scale such that the energy value for $\Sigma 3 (112)$ GB coincides in both methods, one can see that the ADF model overestimates the energy in the range from $0^\circ$ to $70^\circ$ and (to a lesser extent) from $140^\circ$ to $180^\circ$, while the energies in the middle range of angles are underestimated. This appears to be coherent with the discrepancy between the two models in Fig. \ref{missing_atoms_110}. One can conclude that this disagreement is related to the fact that a part of the atoms are situated either more closely one to another or more distantly in the ADF model comparing to configurations of MD. This is due to the changing ``size" of atoms at GBs, already mentioned above, which affects the effective interaction between atomic centers. In any case, the ADF model reproduces very well the two most significant energy cusps: for $\Sigma 3 (112)$ and for $\Sigma 11 (332)$.

\begin{figure}
\center{\includegraphics[angle=-90,width=0.45\textwidth]{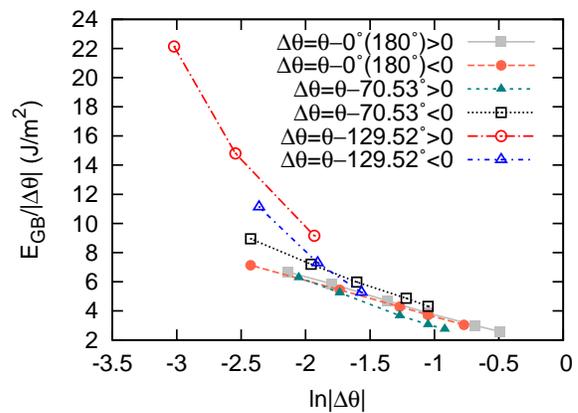}}
\caption{The energy of $[110]$ tilt GBs in a logarithmic scale. The six curves correspond to the six ranges of tilt angles, adjacent to $0^\circ(180^\circ)$, $70.53^\circ$, and $129.52^\circ$. The legend gives how $\Delta\theta$ relates to $\theta$ for each range.}
\label{energy_100_ln}
\end{figure}

\begin{figure*}[t]

\includegraphics[angle=0,width=0.17\textwidth]{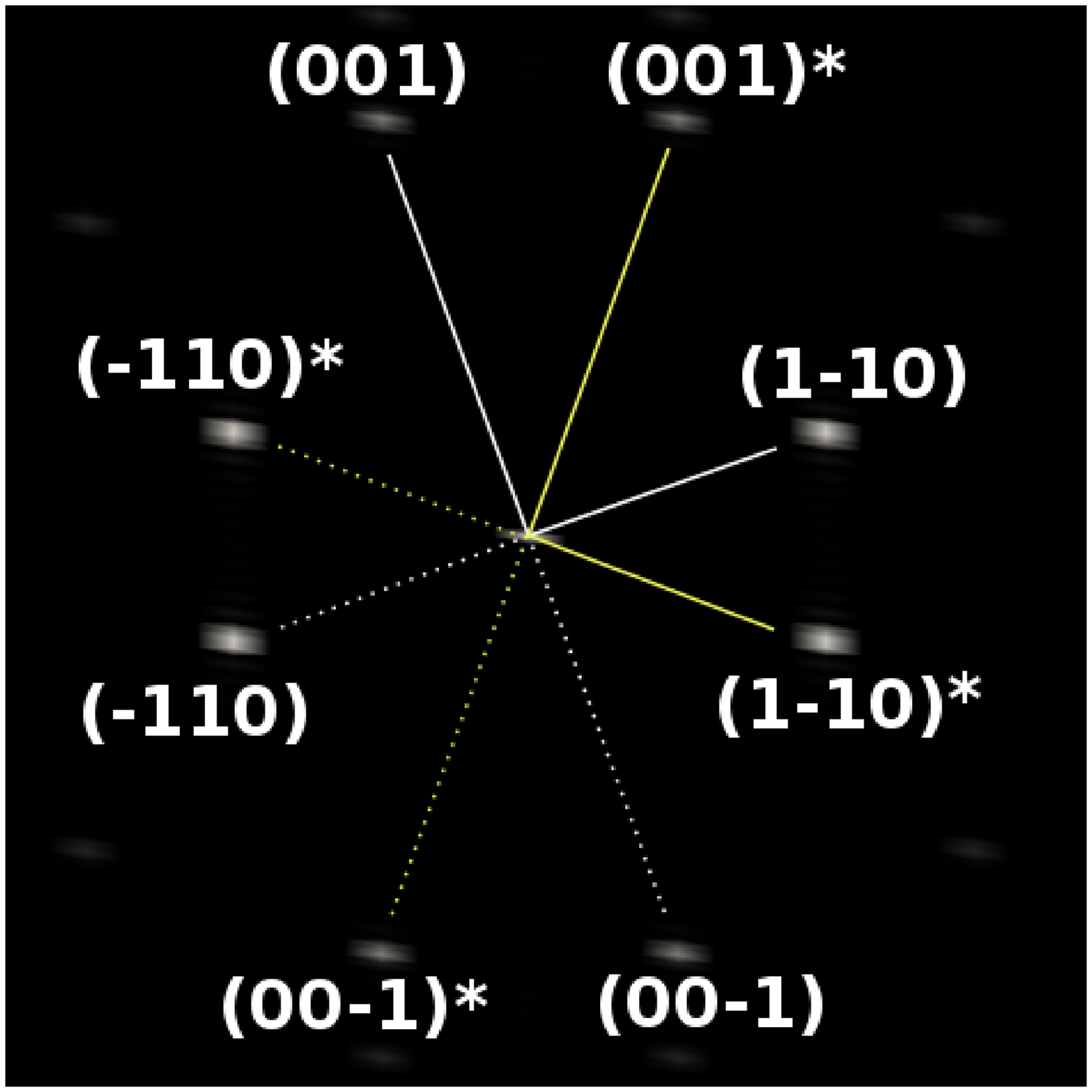}
\hspace{0.2cm}
\includegraphics[angle=0,width=0.17\textwidth]{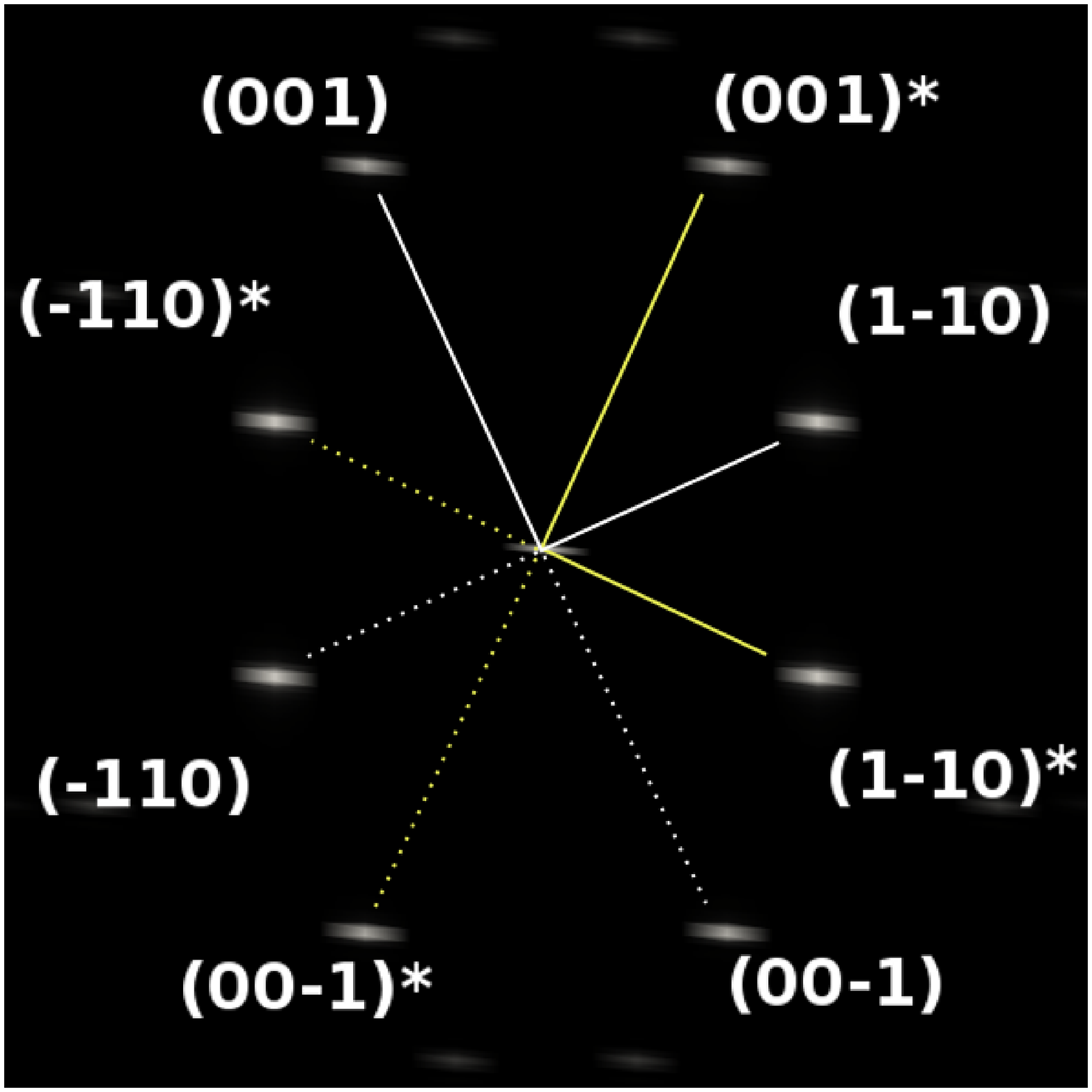}
\hspace{0.2cm}
\includegraphics[angle=0,width=0.17\textwidth]{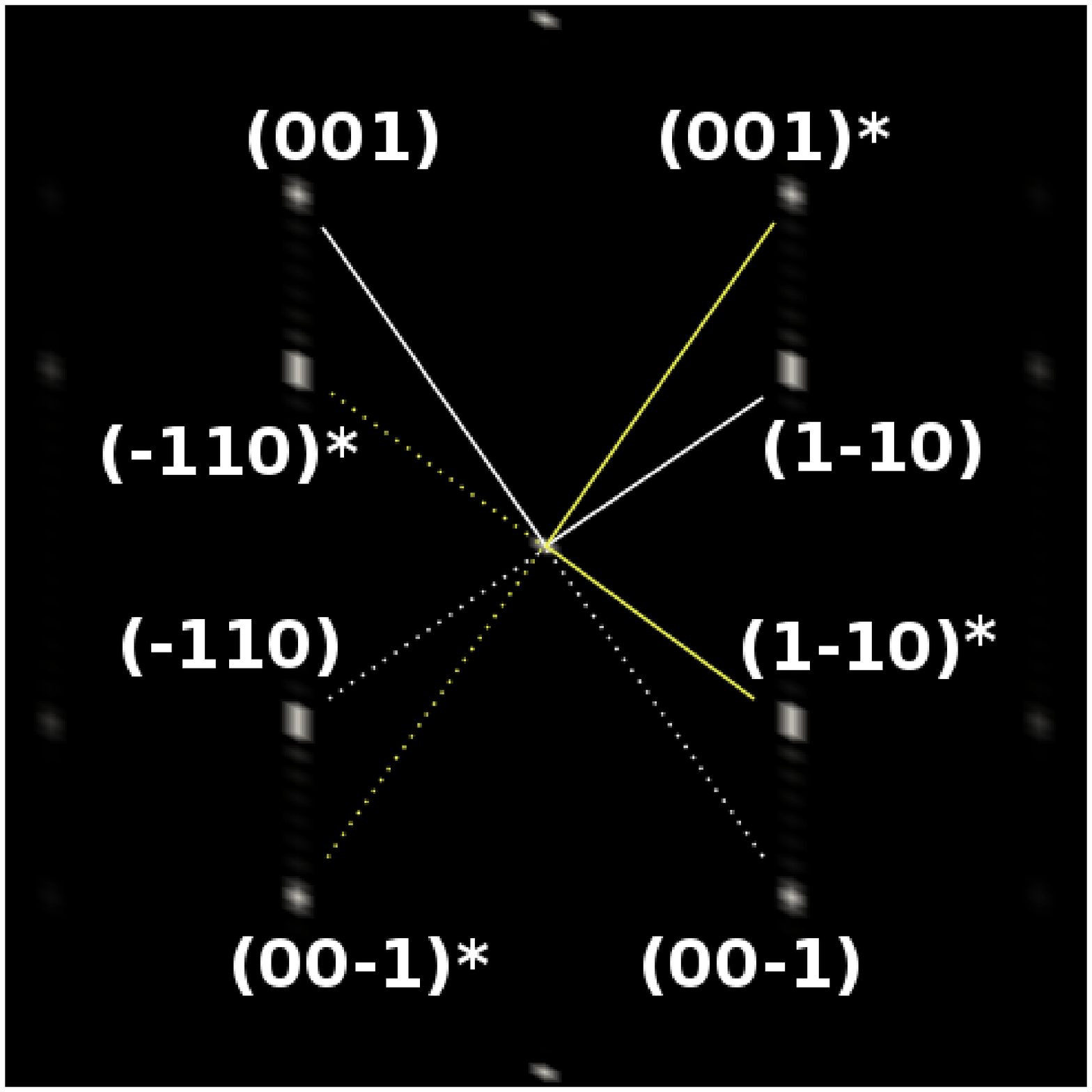}
\hspace{0.2cm}
\includegraphics[angle=0,width=0.17\textwidth]{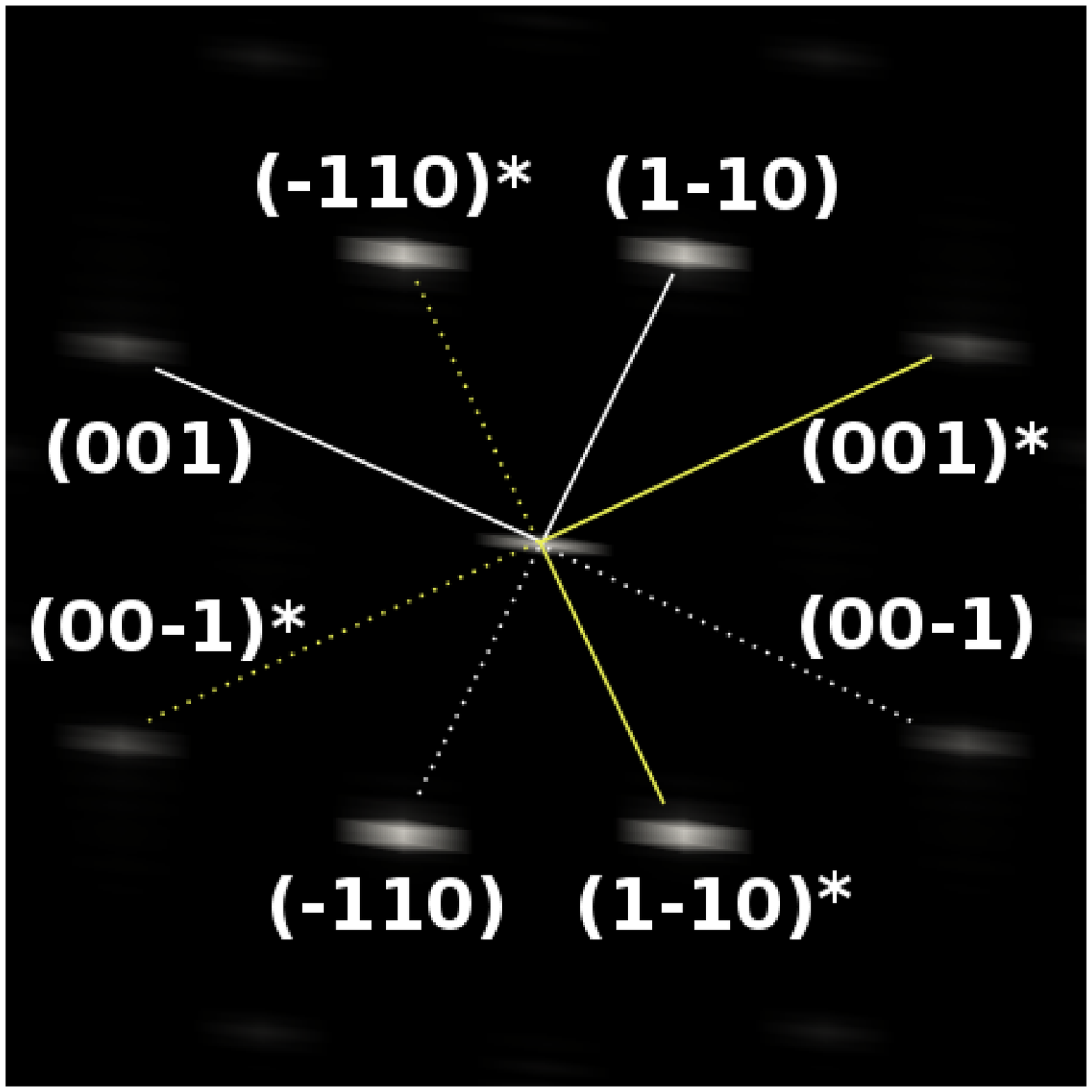}
\hspace{0.2cm}
\includegraphics[angle=0,width=0.17\textwidth]{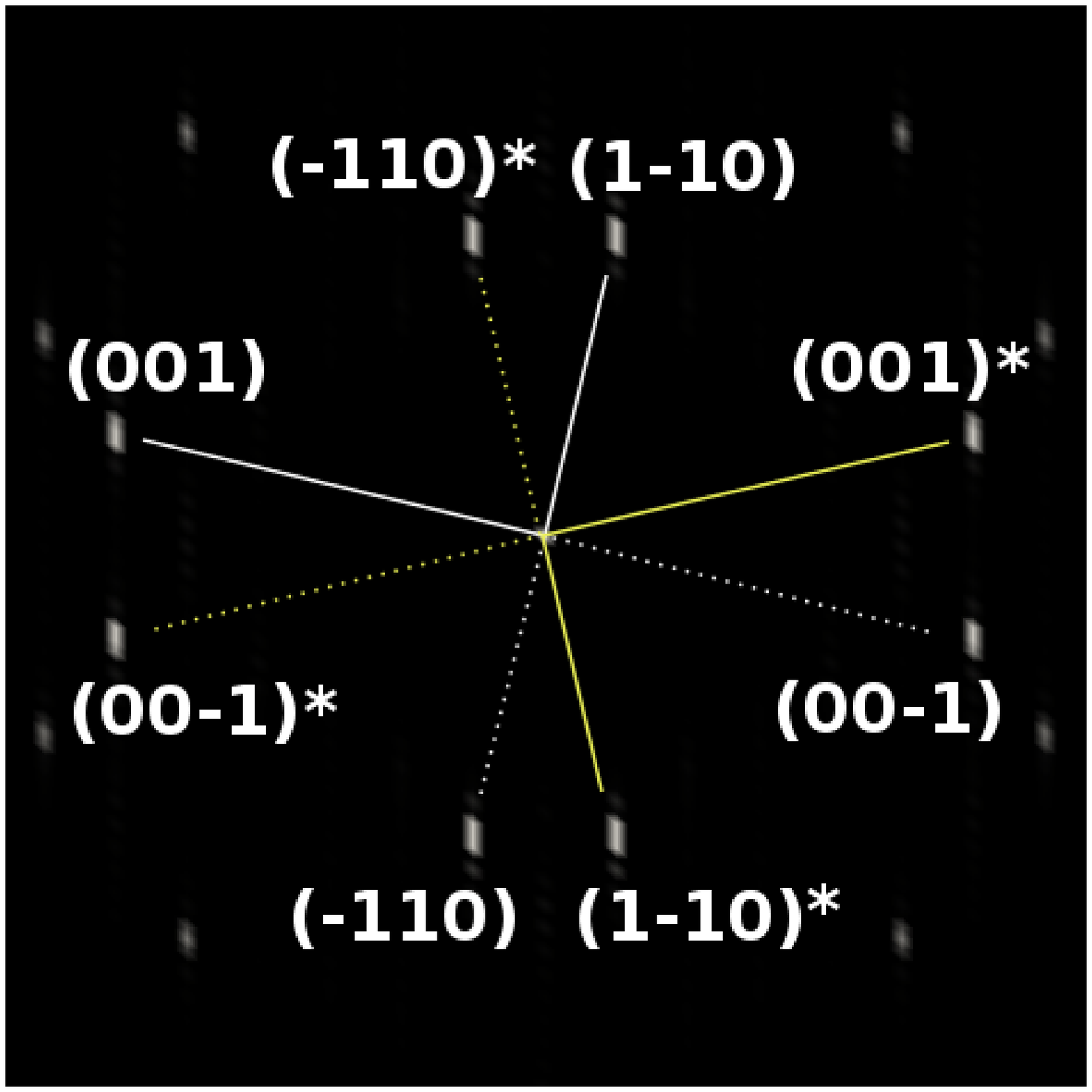}

\includegraphics[angle=0,width=0.082\textwidth]{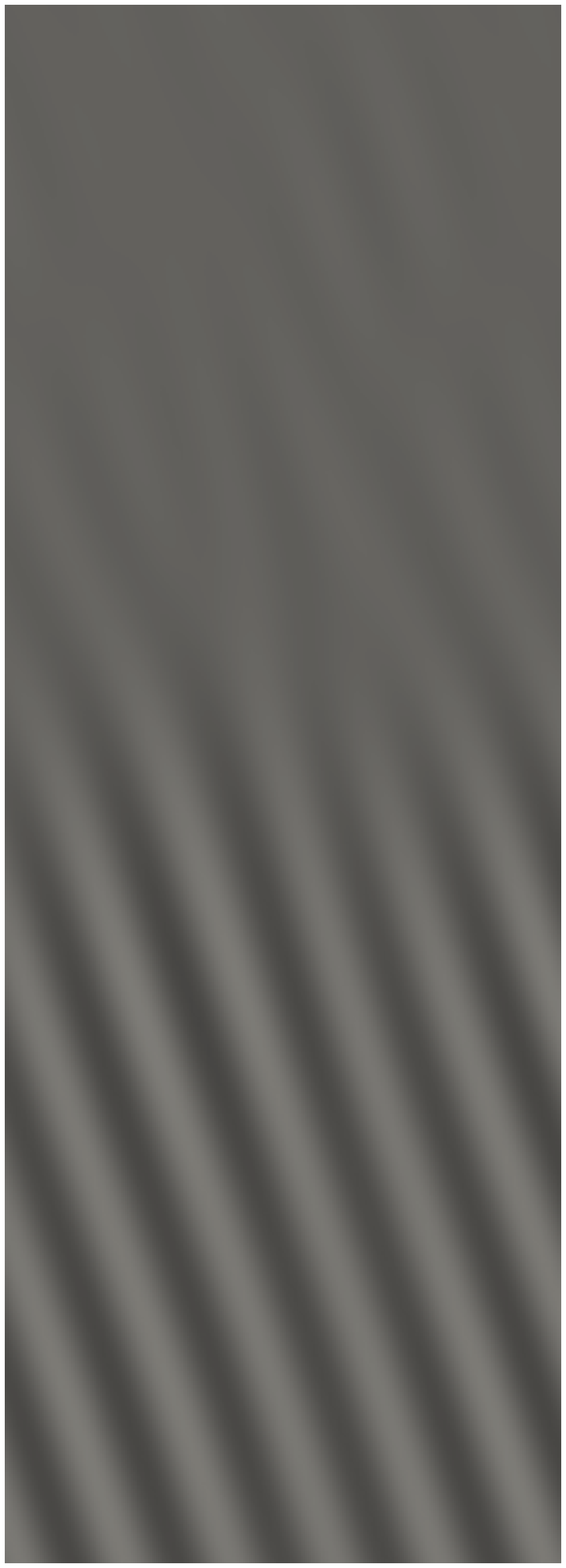}
\includegraphics[angle=0,width=0.082\textwidth]{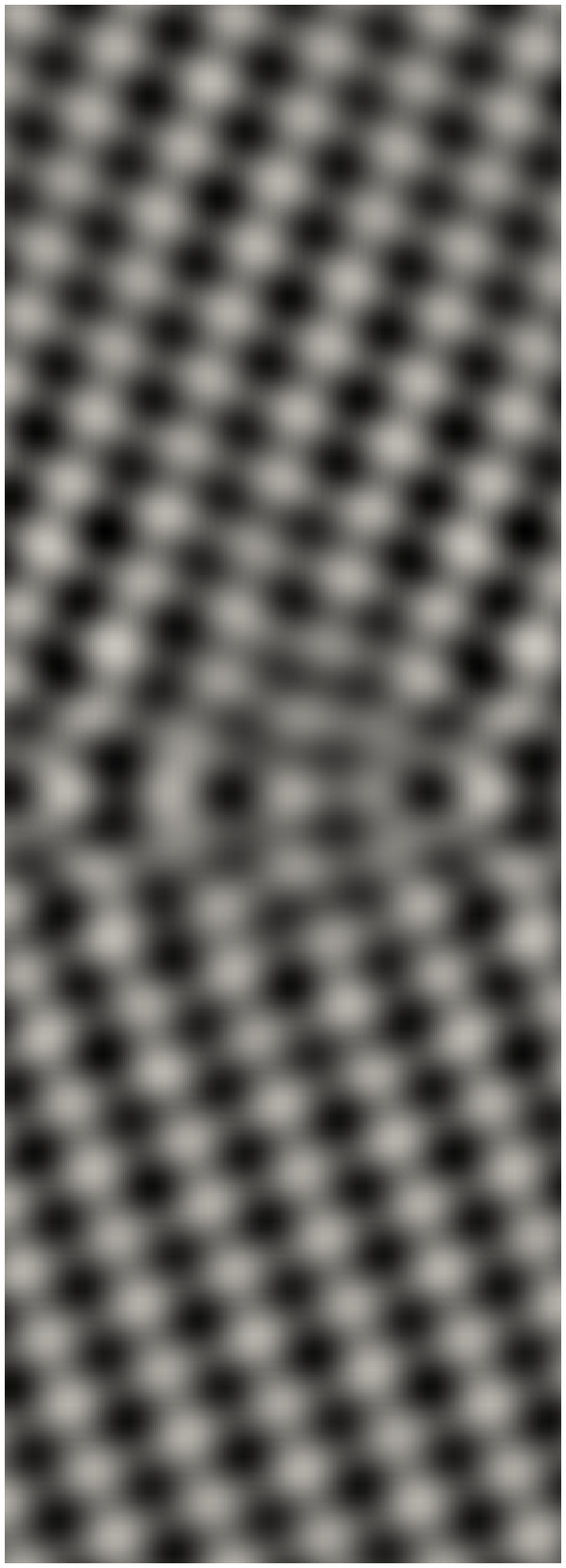}
\hspace{0.2cm}
\includegraphics[angle=0,width=0.082\textwidth]{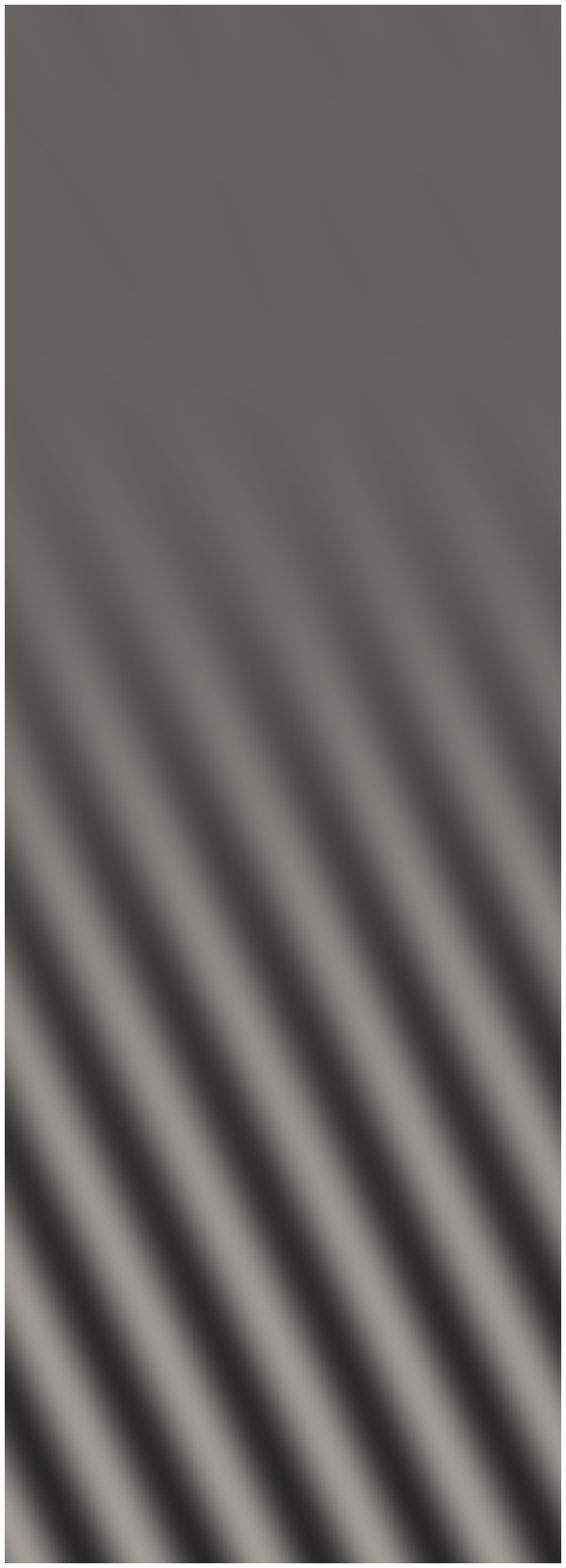}
\includegraphics[angle=0,width=0.082\textwidth]{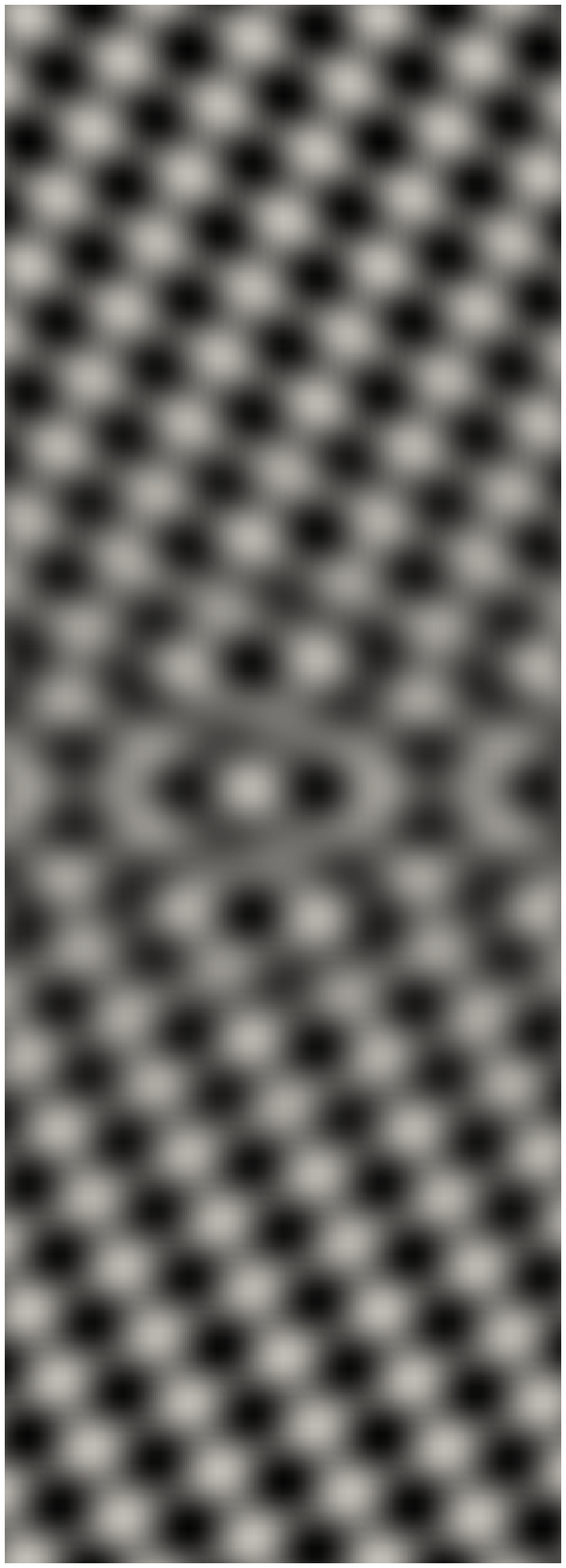}
\hspace{0.2cm}
\includegraphics[angle=0,width=0.082\textwidth]{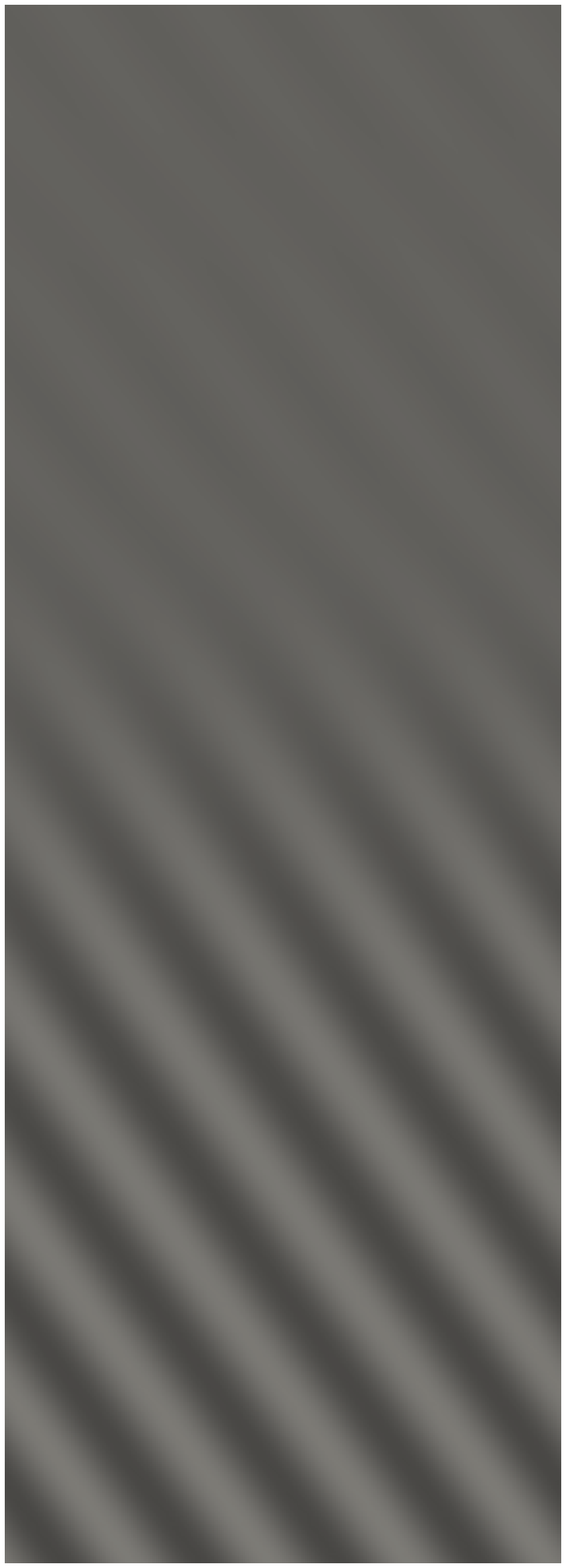}
\includegraphics[angle=0,width=0.082\textwidth]{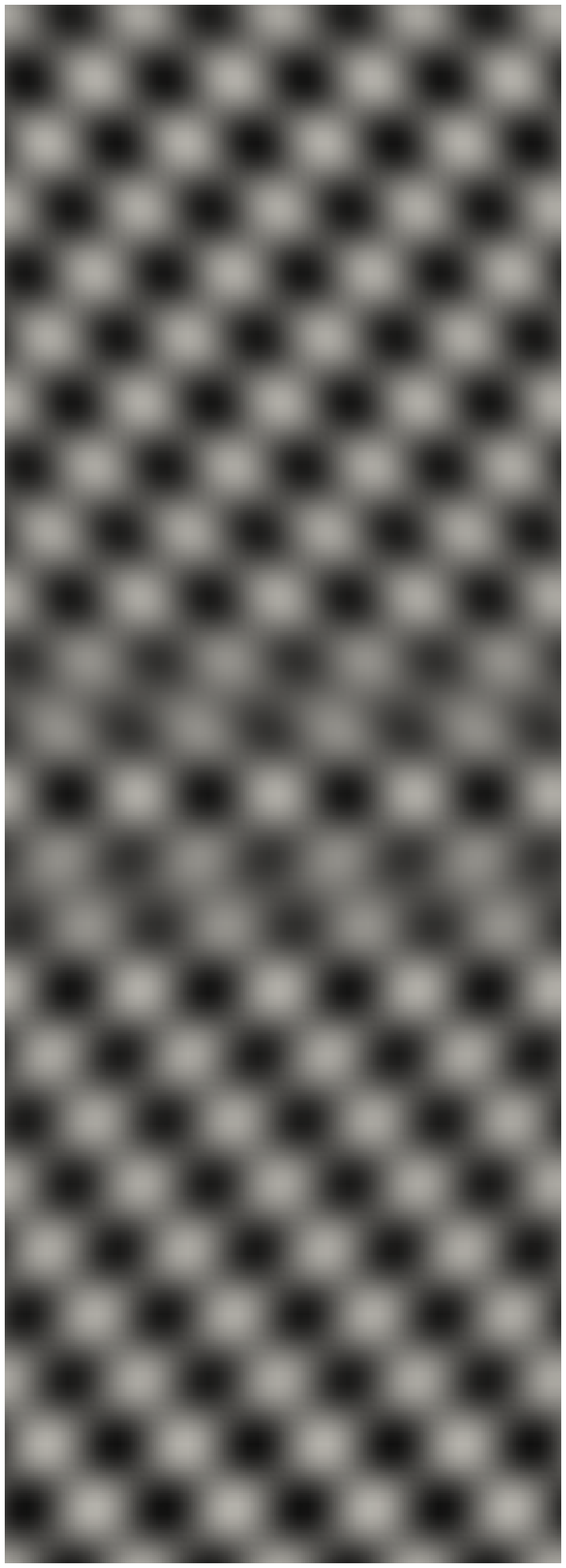}
\hspace{0.2cm}
\includegraphics[angle=0,width=0.082\textwidth]{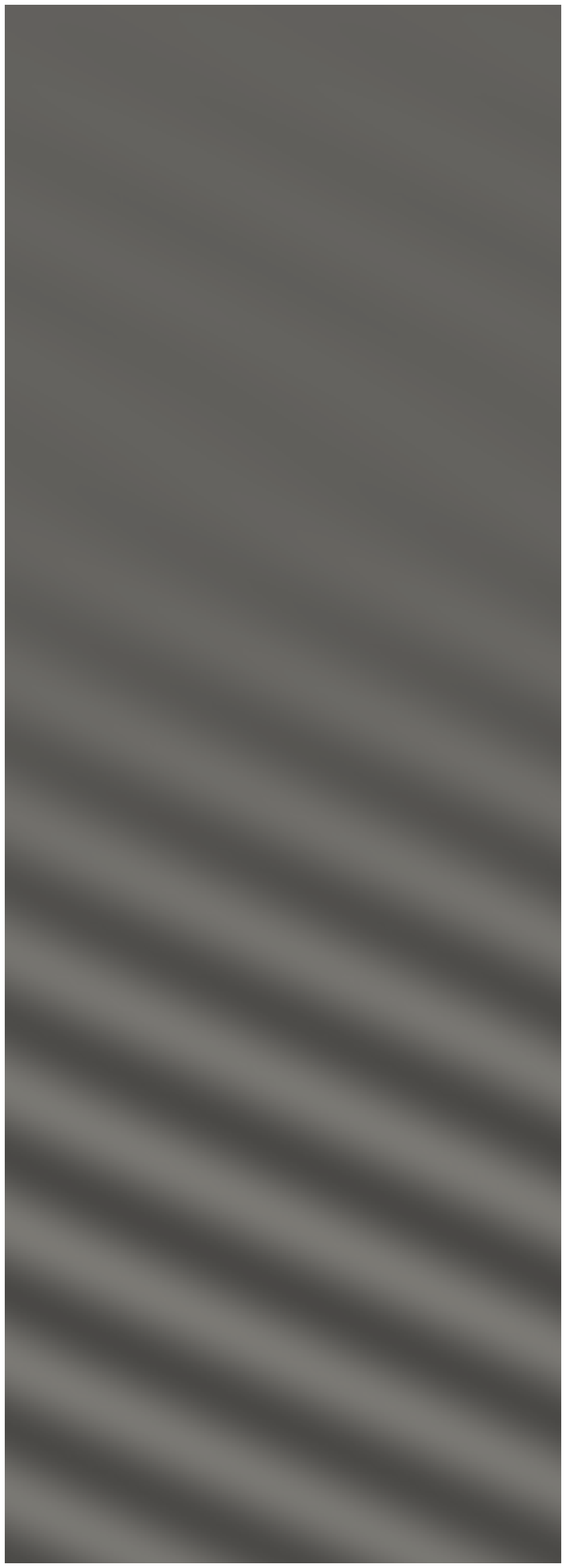}
\includegraphics[angle=0,width=0.082\textwidth]{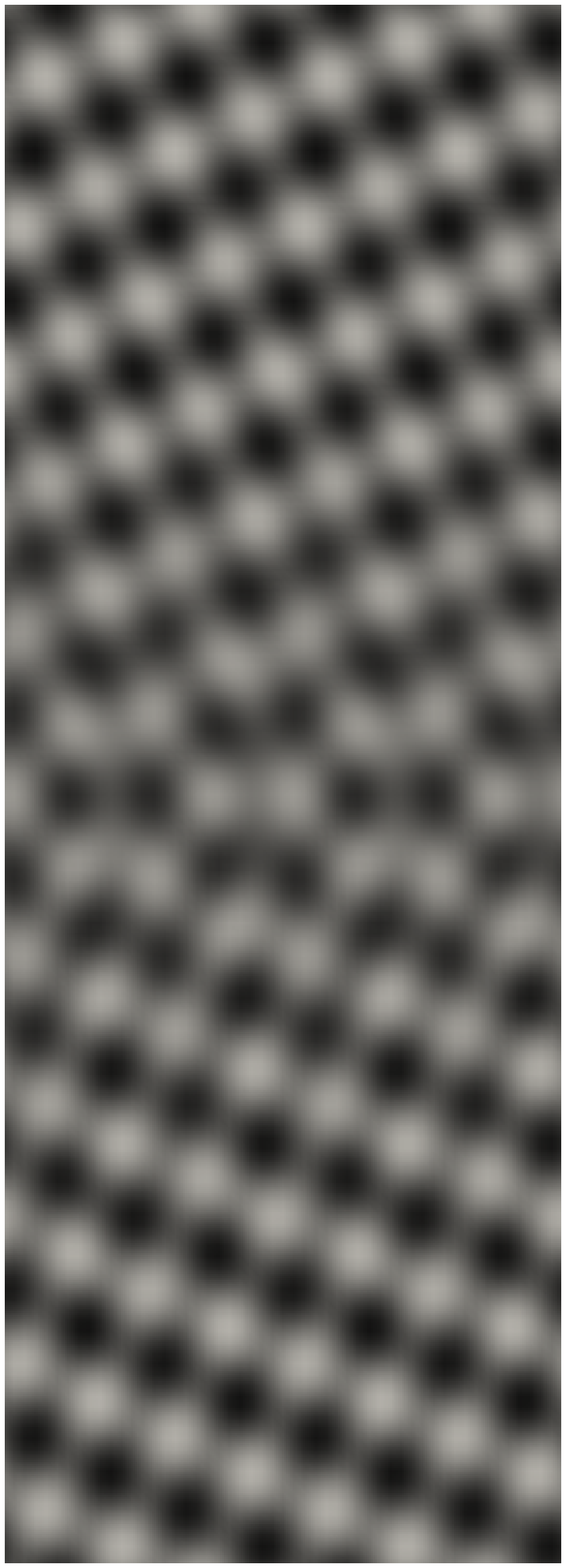}
\hspace{0.2cm}
\includegraphics[angle=0,width=0.082\textwidth]{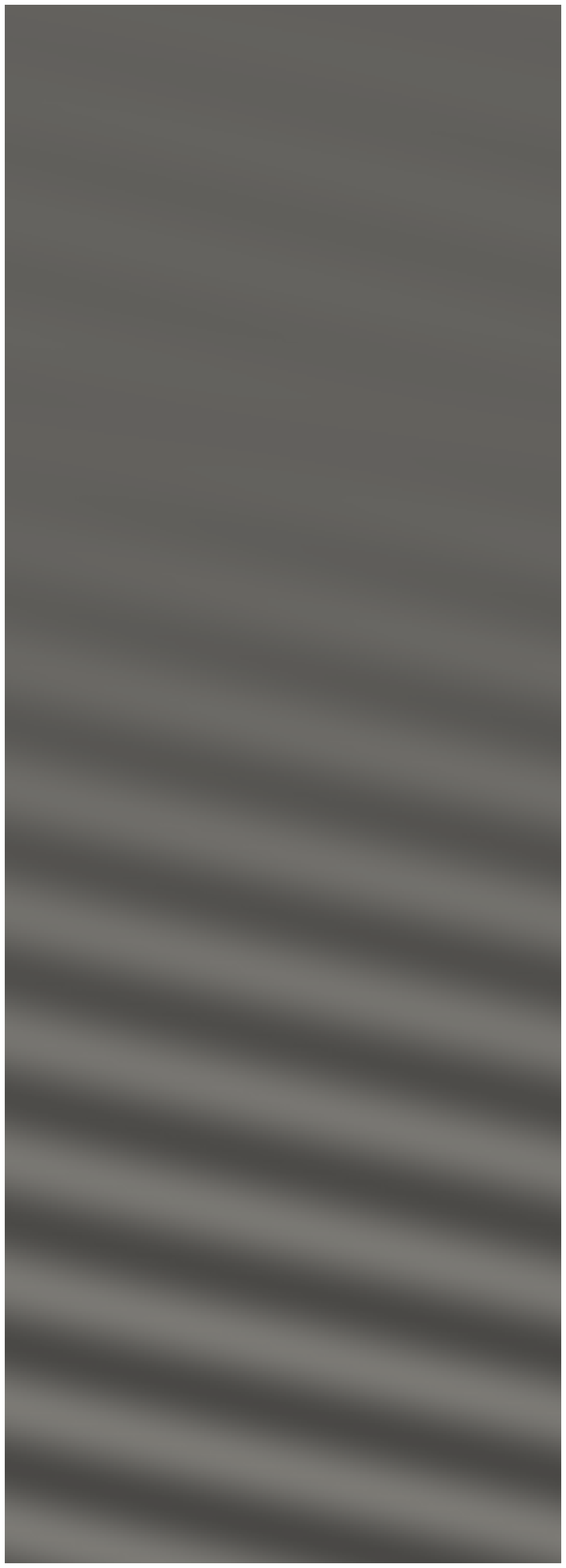}
\includegraphics[angle=0,width=0.082\textwidth]{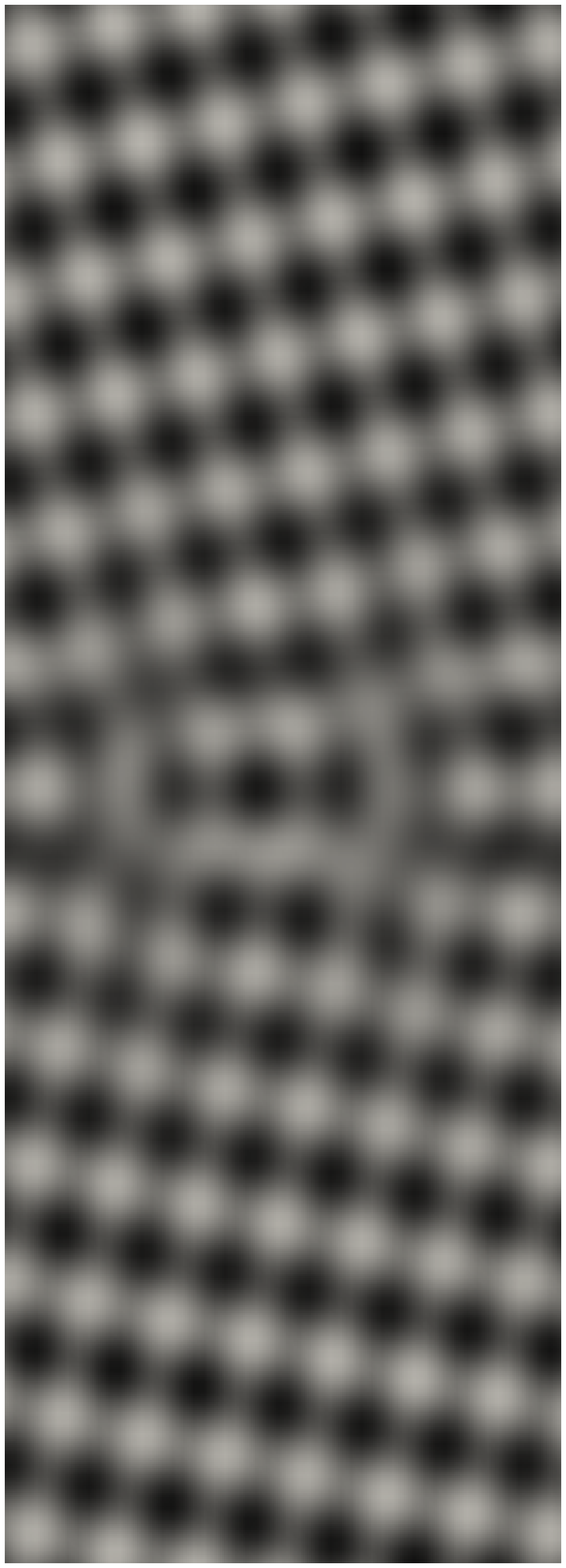}

{\bf (a)} \hspace{2.8cm} {\bf (b)} \hspace{2.8cm} {\bf (c)} \hspace{2.8cm} {\bf (d)} \hspace{2.8cm} {\bf (e)}

\caption{Diffraction patterns (upper row), and real space reconstruction based on the brightest spot $(1\overline{1}0)$ and its conjugate $(\overline{1}10)$ (on the left of each column) or on all the spots indicated in the diffraction patterns (on the right of each column) for a [110] tilt GB of (a) $38.94^\circ$ [$\Sigma 9 (114)$]; (b) $50.48^\circ$ [$\Sigma 11 (113)$]; (c) $70.53^\circ$ [$\Sigma 3 (112)$]; (d) $129.52^\circ$ [$\Sigma 11 (332)$]; (e) $153.47^\circ$, [$\Sigma 19 (221)$].}
\label{HRTEM-110}
\end{figure*}

It is well known that the GB energy behavior is logarithmic at small misorientations as well as in the vicinity of energy cusps. Its evolution can be derived theoretically for low-angle GBs ($< 20^\circ$ and $> 160^\circ$), owing to the continuous elasticity theory and the dislocation model of GBs [\onlinecite{ReadShockley}], as well as for high-angle GBs, using the structural unit model of GBs [\onlinecite{SuttonVitek1983,Gui-JinVitek1986}]. In Fig. \ref{energy_100_ln}, three pairs of curves are presented, each pair corresponding to one of the three ranges of the tilt angle $\theta$: small misorientations ($\sim 0^\circ$ or $\sim 180^\circ$, which are equivalent), the vicinity of the $\Sigma 3 (112)$ and $\Sigma 11 (332)$ cusps. The form in which the energy is presented: $E_{\mathrm{GB}}/|\Delta\theta|$ versus $\ln|\Delta\theta|$, corresponds to that of Read and Shockley [\onlinecite{ReadShockley}]. The deviation $\Delta\theta=\theta - \theta_0$, and, depending on the angles range considered, $\theta_0=0^\circ (180^\circ)$, $70.53^\circ$, and $129.52^\circ$. Linear (or nearly such) dependencies in Fig. \ref{energy_100_ln} confirm once again the validity of the ADF approach to modeling of GB structures.

We have demonstrated that the ADF model reproduces correctly the atomic patterns of all typical symmetrical tilt GBs considered. A numerical tool, such as the ADF model, able to model polycrystalline structures of arbitrary geometry, can be helpful to analysis of HRTEM images. We show in the following that our data can be readily presented in a form analogous to that of HRTEM images.

In Fig. \ref{HRTEM-110}, the upper row presents the distribution of intensity $I(k)\sim \rho_{\bf k}\rho_{-\bf k}$ [see Eq. (\ref{rho_k}) for $\rho_{\bf k}$] in the $k_x=0$ plane of the reciprocal space for some of the bicrystals with [110] tilt GBs modeled above. The intensity maxima correspond to atomic planes that all contain the [110] axis, which is the common axis of the two grains. Reconstruction of the atomic density field $\rho({\bf r})$ from $\rho_{\bf k}$ (by selecting one or several diffraction spots in the plane $k_x=0$ during the inverse Fourier transformation) leads to a set of atomic lattice fringes or to intersections of several sets.

The results of reconstruction based on diffraction spots corresponding to the most dense-packed atomic planes of the lower grain [$(1\overline{1}0)$ and $(\overline{1}10)$ spots] are presented on the left below each diffraction pattern in Fig. \ref{HRTEM-110}. Those obtained using reflections resulting from the two most dense packed planes of both grains, that is, the spots $(1\overline{1}0)$, $(1\overline{1}0)^*$, $(\overline{1}10)$, $(\overline{1}10)^*$, $(001)$, $(001)^*$, $(00\overline{1})$, and $(00\overline{1})^*$, are shown on the right below each diffraction pattern. The atomic ``columns" at this point are merely a result of intersection of several atomic lattice fringes like the ones shown on the left for each orientation.

The left-side reconstructions in Fig. \ref{HRTEM-110} illustrate the capacity of the method to highlight regions with particular orientations of atomic planes, similarly to regular TEM technique. Simply using more diffraction spots during the reconstruction results in images analogous to those of the HRTEM. Indeed, the procedure we use constitutes at the same time the basis of the HRTEM phase contrast imaging. We thus can legitimately expect that our images closely resemble those which could have been obtained by HRTEM, although the latter can be affected by other factors as well.

\section{Conclusions}

In this paper, the ADF model with a logarithmic form of the local part of the free energy functional was applied to model the atomic structure of grain boundaries. The parameters of the model were chosen in order to reproduce the structure factor of iron at the melting point. The ADF model reproduced closely enough the atomic structure of symmetrical tilt GBs in iron (obtained by MD simulation in Ref. [\onlinecite{Tschopp2012}]) as well as the most remarkable features of the energy of high-angle [110] tilt GBs. All grain boundary configurations were eventually relaxed in MD simulation with an EAM potential for iron. It was confirmed that they are close to equilibrium ones. The most notable advantage over the regular PFC model is that atomic positions are clearly localized for any GB geometry. The two most significant energy cusps of $[110]$ tilt GBs, that is $\Sigma 3 (112)$ and $\Sigma 11 (332)$, have been reproduced, with the energy decreasing logarithmically in their vicinity, as expected. The model is thus claimed to be a powerful tool for construction and study of large disorientation grain boundaries, which are encountered most frequently in experimentally analyzed materials.

It is important to mention once again that the application of the ADF method is not at all limited to modeling of GB atomic configurations but can cover various GBs related phenomena evolving on diffusive time scales. The equilibrium segregation can be studied by introducing a second atomic density field for impurity/solute atoms. Inclusion of atomic vacancies and interstitials into the consideration will permit to study non equilibrium segregation. The Fourier transform of the ADF gives a possibility to build images that can be compared directly to the HRTEM.

\section{Acknowledgments}

The authors are grateful to A. G. Khachaturyan for fruitful discussions. This research is supported by Carnot and the simulations have been performed at the Centre de Ressources Informatiques de Haute-Normandie (CRIHAN) under Project No. 2012008. We also acknowledge the partial financial support from Material Ageing Institute - Scientific Network (MAI-SN). This work is a part of the research program of the EDF-CNRS joint laboratory EM2VM (Study and Modeling of the Microstructure for Ageing of Materials). O. K. is grateful to T. Platini for his comments on the manuscript.

\nocite{*}

\bibliography{biblio}

\end{document}